\documentclass[11pt]{amsart}
\usepackage{amssymb,mathrsfs,graphicx}
\newcommand{\bx}{{\bf x}}
\newcommand{\by}{{\bf y}}

\newcommand{\bu}{{\bf u}}
\newcommand{\bv}{{\bf v}}
\newcommand{\bw}{{\bf w}}

\def\cstok#1{\leavevmode\thinspace\hbox{\vrule\vtop{\vbox{\hrule\kern1.25pt
\hbox{\vphantom{\tt/}\thinspace{\tt#1}\thinspace}}
\kern1.25pt\hrule}\vrule}\thinspace}

\def\cstok#1{\leavevmode\thinspace\hbox{\vrule\vtop{\vbox{\hrule\kern3pt
\hbox{\vphantom{\tt/}\thinspace{\tt#1}\thinspace}}
\kern3pt\hrule}\vrule}\thinspace}

\title{APPENDIX to paper: A geometric level-set formulation of a plasma-sheath interface}

\author{Mikhail Feldman}
\address{Department of Mathematics, University of Wisconsin-Madison, USA}
\email{feldman@math.wisc.edu}

\author{Seung-Yeal Ha}
\address{Department of Mathematics, Seoul National University, KOREA}
\email{syha@math.snu.ac.kr}

\author{Marshall Slemrod}
\address{Department of Mathematics, University of Wisconsin-Madison, USA}
\email{slemrod@math.wisc.edu}

\begin{document}

\newtheorem{theorem}{Theorem}[section]
\newtheorem{lemma}{Lemma}[section]
\newtheorem{proposition}{Proposition}[section]
\newtheorem{remark}{Remark}[section]
\newtheorem{definition}{Definition}[section]

\renewcommand{\theequation}{\thesection.\arabic{equation}}
\renewcommand{\thetheorem}{\thesection.\arabic{theorem}}
\renewcommand{\thelemma}{\thesection.\arabic{lemma}}
\newcommand{\bbr}{\mathbb R}
\def\charf {\mbox{{\text 1}\kern-.24em {\text l}}}

\begin{abstract}
In this paper, we present appendices employed in the paper "A
geometric level-set formulation of a plasma-sheath interface" by
the authors.
\end{abstract}
\maketitle


\appendix

%
%
\begin{section}{{\bf Local existence of sheath solutions}}
In this appendix, we present a series of a priori estimates for
the approximate solutions constructed in {\it Step 0} - {\it Step
7} in Section 7 and then give the proof Theorem 7.3.

\begin{subsection}{{\bf Basic a priori estimates}} In this part,
 we give {\it a priori} estimates for the approximate solutions constructed in
{\it Step 0} - {\it Step 7}.


\begin{subsubsection}{{\bf A priori estimates for ${\it Step 1}$}} In Lemmas A1-A3,
we will give a proof of the existence, uniqueness and regularity
for $n$ as given in {\it Step 1} of Section 7.2. \newline

We first consider the equation for a characteristic curve.  For
given $(\bx,t)$,
\begin{equation}
\displaystyle   \partial_{s} \mbox{\boldmath $\chi$}(s,t,\bx)  =
\bv (\mbox{\boldmath $\chi$}(s,t,\bx), s), \quad  \mbox{\boldmath
$\chi$}(t,t,\bx) = \bx, \qquad 0 \leq s \leq T. \label{char}
\end{equation}
In what follows,  we will use calculus type estimates for the
H\"older seminorm.
 For $f_i  \in C^{0,\gamma}(\bar \Lambda_s(T;\bv))~~i=1,2,$ we have
\begin{eqnarray}
&& \displaystyle [[f_1f_2]]_{0,\gamma} \leq  [[f_1]]_{0,\gamma}
|||f_2|||_{0}
+ |||f_1|||_{0} [[f_2]]_{0,\gamma}, \label{H1} \\
&& \displaystyle [[e^f]]_{0,\gamma} \leq e^{|||f|||_{0}}
[[f]]_{0,\gamma}. \label{H2}
\end{eqnarray}
Here $[[\cdot]]_{0,\gamma}$ and $|||\cdot|||_{0}$ denote the
H\"older and $esssup$ norms defined on the same space-time region.
\newline

In the following Lemma, we use simplified notation for balls in
$\bbr^2$:
\[ B_1:= B(0,r_b + 3K_0 \delta^* T_0) \qquad \mbox{ and } \qquad B_2:= B(0,r_b + 6K_0 \delta^* T_0). \]
%
%
\begin{lemma}
There exists a sufficiently small constant $T_0 >0$ and a unique
solution  $\mbox{\boldmath $\chi$}$
 to the equation (\ref{char}) satisfying the following estimates:
For $0 < T \leq T_0,~~\bv \in {\mathcal B}(T)$,
\begin{enumerate}
\item The forward characteristic curve $\mbox{\boldmath
$\chi$}(s,0,\bx),~s \geq 0, \quad \bx  \in \Omega_s^1(0;\bv)
\subset (B_1 - \Omega_0)$ hits the target boundary $\partial
\Omega_0$ and the ion-density in the region $\Lambda_s^1(T;\bv)$
is given by
\[
n(\mbox{\boldmath $\chi$}(t,0,\bx),t) = n_0(\bx)
\exp\Big(-\int_0^t (\nabla \cdot \bv)(\mbox{\boldmath $\chi$}(s,
0, \bx), s)ds \Big), \quad \bx \in \Omega_s^1(0;\bv).
\]
\item $\mbox{\boldmath $\chi$}(s,t,\bx) \in C^{1,\gamma}([0,T]
\times [0,T] \times \bbr^2)$ and $\quad \displaystyle
\sup_{s,t,\bx} \max_{i,j=1,2} |\partial_{x_j} \chi^i| \leq 2.$
\item Suppose that $\bv_i \to \bv$ in $C^{1,\gamma}(\bar
\Lambda(T))$ and let $\mbox{\boldmath $\chi$}_i$ and
$\mbox{\boldmath $\chi$}$ be the characteristic curves
corresponding to $\bv_i$ and $\bv$ respectively. Then for $(\bx,t)
\in \Lambda_s^1(T;\bv)$,
\[ \mbox{\boldmath $\chi$}_i(\cdot,t,\bx) \to  \mbox{\boldmath $\chi$}(\cdot,t,\bx) \quad \mbox{ in } C^{1,\gamma}([0,T]). \]
\item $\mbox{\boldmath $\alpha$}(\bx,t) :=\mbox{\boldmath
$\chi$}(0,t,\bx)$ is Lipschitz continuous in $(\bx,t) \in
\Lambda(T)$ with a Lipschitz constant 4, i.e.
\[ |\mbox{\boldmath $\alpha$}(\bx,t) - \mbox{\boldmath $\alpha$}(\by,s)| \leq 4|(\bx,t) - (\by,s)|. \]
\end{enumerate}
\end{lemma}
\begin{proof}
\noindent (i) It follows from the dissipative condition
$({\mathcal D}2)$ in the definition of ${\mathcal B}(T)$, we have
\[ \bv(\bx,t) \cdot \bx \leq -\frac{\eta_0}{2} |\bx|^2, \quad (\bx, t) \in (B_2 - \Omega_0) \times [0, T]. \]
Then we have for $(\bx, t) \in (B_2 - \Omega_0) \times [0, T]$,
\[
\frac{d}{ds}|\mbox{\boldmath $\chi$}(s,t,\bx)|^2 = 2\langle
\bv(\mbox{\boldmath $\chi$}(s,t,\bx),s), \mbox{\boldmath
$\chi$}(s,t,\bx) \rangle \leq -\eta_0 |\mbox{\boldmath
$\chi$}(s,t,\bx)|^2.
\]
Here $\langle \cdot,\cdot \rangle$ denotes the standard inner
product in $\bbr^2$. Hence the characteristic $\mbox{\boldmath
$\chi$}(s,t,\bx)$ satisfies
\[ |\mbox{\boldmath $\chi$}(s,0,\bx)| \leq e^{-\frac{\eta_0 s}{2}} |\mbox{\boldmath $\chi$}(0,0,\bx)| = e^{-\frac{\eta_0 s}{2}} |\bx|. \]
So $\mbox{\boldmath $\chi$}(s,0,\bx)$ has decreasing magnitude and
must hit the target at some positive $s$. \newline

Let $T \leq T_0$ and we define the subregions $\Lambda_s^1(T;\bv),
\Omega_s^1(T;\bv)$ of $\Lambda(T)$ and $\Omega_s(0)$
 as in {\it Step 1} of Section 7.2.1. Then the characteristic curve
 $\mbox{\boldmath $\chi$}(s,0,\bx), \quad (\bx,0)  \in \Omega_s^1(0) \times \{ t =0 \}$ hits the target boundary $\partial \Omega_0$
 and will provide the ion density $n$ at the target boundary, i.e.,
\[
n(\mbox{\boldmath $\chi$}(t,0,\bx),t) = n_0(\bx)
\exp\Big(-\int_0^t (\nabla \cdot \bv)(\mbox{\boldmath $\chi$}(s,
0, \bx), s)ds \Big), \quad \bx \in \Omega_s^1(0;\bv).
\]

\begin{figure}[!hbtp]
\includegraphics{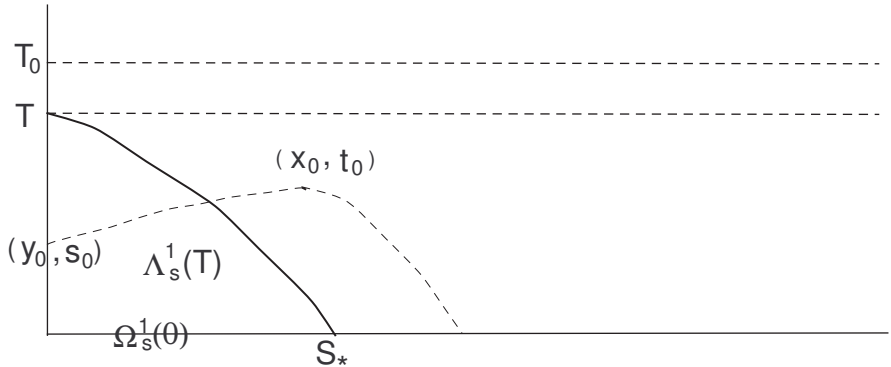}
\caption{Schematic diagram of the geometry of characteristic
curves}
\end{figure}

\begin{remark}
We briefly summarize the geometry of characteristic curves in the
space-time region $\Lambda(T)$ (see Figure 5). \newline

The region $\Lambda_s^1(T)$ will be completely covered by the
characteristic curves $\mbox{\boldmath $\chi$}(s,0,\bx), (\bx,0)
\in \Omega_s^1(0;\bv) \times \{ t = 0\}$ and they are pointing
toward the target for positive $s$. On the other hand, all
backward characteristic curves $\mbox{\boldmath $\chi$}(s,t,\bx),
0 \leq s \leq t, (\bx,t) \in \Lambda(T)- \Lambda_s^1(T)$ will
either hit the initial region $ (B(0,3\delta^*) - \Omega_0) \times
\{ t =0 \}$ at $s=0$ or the target boundary $\partial \Omega_0$ at
some $s \in [0,t)$ (see Figure 5). However the latter situation
will not happen, for example, suppose the backward characteristic
curve $\mbox{\boldmath $\chi$}(s,t_0,\bx_0),  0 \leq s \leq t_0,
(\bx_0,t_0) \in \Lambda(T)- \Lambda_s^1(T)$ hits the target
boundary at $s = s_0$ at $\by_0$:
\[ \by_0 := \mbox{\boldmath $\chi$}(s_0,t_0,\bx_0). \]
Then forward characteristic curve $\mbox{\boldmath
$\chi$}(s,s_0,\by_0), s \in [s_0, t_0]$ will have the same image
of a  trajectory as $\mbox{\boldmath $\chi$}(s,t_0,\bx_0), s \in
[s_0, t_0], (\bx_0,t_0) \in \Lambda(T)- \Lambda_s^1(T)$, but this
is impossible since by the strong dissipation assumption
${\mathcal D}2$ in the Definition 7.1, no forward characteristic
curves will be issued from the target boundary.
\end{remark}

\noindent (ii) The first part of the proof for (1) follows from
the standard theory of ordinary differential equations. In fact we
gain regularity in the $s$-variable, i.e.,
\[  \mbox{\boldmath $\chi$}(\cdot,t,\bx) \in C^{2,\gamma}([0,T]). \]

We differentiate (\ref{char}) with respect to $x_j$ to get
\[
\begin{cases}
 \partial_{s} \partial_{x_j} \chi^k(s,t,\bx)
 = \nabla v_k (\mbox{\boldmath $\chi$}(s,t,\bx), s) \cdot \partial_{x_j} \mbox{\boldmath $\chi$}(s,t,\bx), \quad 0 \leq s \leq T \quad k,j \in \{1,2\}, \cr
 \partial_{x_j} \chi^k(t,t,\bx) = \delta_{jk}.
\end{cases}
 \]
Here $\delta_{jk}$ is a Kronecker delta function and $\chi^{k}$ is
the $k$-th component of $\mbox{\boldmath $\chi$}$, k=1,2. \newline

We integrate the above equation along the characteristic curve
$\mbox{\boldmath $\chi$}$ to see
\[ \partial_{x_j} \chi^k(\xi,t,\bx)
= \delta_{jk} - \int_{\xi}^t \nabla v_k (\mbox{\boldmath
$\chi$}(s,t,\bx), s) \cdot \partial_{x_j} \mbox{\boldmath
$\chi$}(s,t,\bx)ds. \] The above relation implies
\[ \sup_{s,t,\bx} \max_{k,j=1,2} |\partial_{x_j} \chi^k| \leq 1 + 6K_0 \delta^* (t-\xi) \sup_{s,t,\bx} \max_{k,j=1,2}  |\partial_{x_j} \chi^k|. \]
Since $t-\xi \leq T \ll 1$, we have
\[ \sup_{s,t,\bx} \max_{k,j=1,2} |\partial_{x_j} \chi^k| \leq 2. \]

(iii) Consider the equations for $\mbox{\boldmath $\chi$}_i$ and
$\mbox{\boldmath $\chi$}$: For $(\bx,t) \in \Lambda_s^1(T;\bv)$,
\[
\begin{cases}
\partial_{\xi} \mbox{\boldmath $\chi$}_i(\xi,t,\bx)  = \bv_i (\mbox{\boldmath $\chi$}_i(\xi,t,\bx), \xi), \cr
\mbox{\boldmath $\chi$}_i(t,t,\bx) = \bx,
\end{cases}
\mbox{ and } \quad
\begin{cases}
\partial_{\xi} \mbox{\boldmath $\chi$}(\xi,t,\bx)  = \bv(\mbox{\boldmath $\chi$}(\xi,t,\bx), \xi), \cr
\mbox{\boldmath $\chi$}(t,t,\bx) = \bx.
\end{cases}
\]
We use the above equations to calculate $\mbox{\boldmath
$\chi$}_i(\xi,t,\bx) - \mbox{\boldmath $\chi$}(\xi,t,\bx)$, and
integrate in $\xi$ from $\xi = s$ to $\xi=t$ to get
\begin{eqnarray}
 \mbox{\boldmath $\chi$}_i(s,t,\bx) - \mbox{\boldmath $\chi$}(s,t,\bx)
 &=& - \int_s^t \Big( \bv_i(\mbox{\boldmath $\chi$}_i(\xi,t,\bx),\xi) -
\bv(\mbox{\boldmath $\chi$}(\xi,t,\bx),\xi) \Big)d\xi \nonumber \\
 &=& -\int_s^t \Big( \bv_i(\mbox{\boldmath $\chi$}_i(\xi,t,\bx),\xi) - \bv_i(\mbox{\boldmath $\chi$}(\xi,t,\bx),\xi) \Big)d\xi \nonumber \\
 &-& \int_s^t \Big(\bv_i(\mbox{\boldmath $\chi$}(\xi,t,\bx),\xi) - \bv(\mbox{\boldmath $\chi$}(\xi,t,\bx),\xi) \Big)d\xi. \nonumber \\
\label{L7.4-1}
\end{eqnarray}
Here we used $\mbox{\boldmath $\chi$}_i(t,t,\bx) = \mbox{\boldmath
$\chi$}(t,t,\bx) = \bx$ and note that
\begin{eqnarray}
&& \int_s^t \Big( \bv_i(\mbox{\boldmath $\chi$}_i(\xi,t,\bx),\xi) - \bv_i(\mbox{\boldmath $\chi$}(\xi,t,\bx),\xi) \Big)d\xi \nonumber \\
&& = \int_s^t \int_0^1 \partial_{s_1}  \bv_i\Big (\mbox{\boldmath
$\chi$}(\xi,t,\bx) + s_1(\mbox{\boldmath $\chi$}_i(\xi,t,\bx) -
 \mbox{\boldmath $\chi$}(\xi,t,\bx),\xi\Big )ds_1 d\xi \nonumber \\
&& = \int_s^t \int_0^1 \nabla_{\bx} \bv_i \Big (\mbox{\boldmath
$\chi$}(\xi,t,\bx) + s_1(\mbox{\boldmath $\chi$}_i(\xi,t,\bx) -
 \mbox{\boldmath $\chi$}(\xi,t,\bx),\xi\Big ) \cdot \Big( \mbox{\boldmath $\chi$}_i(\xi,t,\bx) -
 \mbox{\boldmath $\chi$}(\xi,t,\bx) \Big)ds_1 d\xi. \nonumber \\
 \label{L7.4-1-1}
\end{eqnarray}
We now take the $\bbr^2$-norm  in (\ref{L7.4-1}) and use
(\ref{L7.4-1-1}) to see
\begin{eqnarray*}
&& |\mbox{\boldmath $\chi$}_i(s,t,\bx) - \mbox{\boldmath
$\chi$}(s,t,\bx)| \cr && \hspace{2cm} \leq |||\nabla \bv_i|||_{0,
\bar \Lambda(T)} \int_s^t |\mbox{\boldmath $\chi$}_i(\xi,t,\bx) -
\mbox{\boldmath $\chi$}(\xi,t,\bx)|d\xi + |||\bv_i - \bv|||_{0,
\bar \Lambda(T)}(t-s).
\end{eqnarray*}
Note that Gronwall's inequality yields
\begin{eqnarray}
&& |\mbox{\boldmath $\chi$}_i(s,t,\bx) - \mbox{\boldmath $\chi$}(s,t,\bx)| \nonumber \\
&& \hspace{2cm} \leq |||\bv_i - \bv|||_{0, \bar \Lambda(T)}(t-s)
\Big( 1 + |||\nabla \bv_i|||_{0, \bar \Lambda(T)} (t-s) e^{|||\nabla \bv_i|||_{0,\bar \Lambda(T)} (t-s)} \Big). \nonumber \\
\label{L7.4-1-2}
\end{eqnarray}
By hypothesis (2) of this lemma, we have $\bv_i \to \bv $ in
$C^{1,\gamma}(\bar \Lambda(T))$ as $t \to \infty,$ and this
implies from (\ref{L7.4-1-2}) that
\begin{equation}
||\mbox{\boldmath $\chi$}_i(s,t,\bx) - \mbox{\boldmath
$\chi$}(s,t,\bx)||_{0,[0,T]} \to 0 \quad \mbox{ as } i \to \infty.
\label{L7.4-2}
\end{equation}
Next we show
\begin{equation}
|| \partial_s \mbox{\boldmath $\chi$}_i(\cdot,t,\bx) - \partial_s
\mbox{\boldmath $\chi$}(\cdot,t,\bx)||_{0,[0,T]} \to 0\quad \mbox{
as } i \to \infty. \label{L7.4-2-1}
\end{equation}
Note that (\ref{char}) implies
\begin{eqnarray*}
&& ||\partial_s \mbox{\boldmath $\chi$}_i(\cdot ,t,\bx) -
\partial_s \mbox{\boldmath $\chi$}(\cdot,t,\bx)||_{0,[0,T]} \cr &&
\leq || \bv_i(\mbox{\boldmath $\chi$}_i(\cdot,t,\bx),\cdot) -
\bv_i(\mbox{\boldmath $\chi$}(\cdot,t,\bx),\cdot)||_{0,[0,T]} + ||
\bv_i(\mbox{\boldmath $\chi$}(\cdot,t,\bx),\cdot) -
\bv(\mbox{\boldmath $\chi$}(\cdot,t,\bx),\cdot)||_{0,[0,T]} \cr &&
\leq |||\nabla \bv_i|||_{0, \bar \Lambda(T)} || \mbox{\boldmath
$\chi$}_i(\cdot,t,\bx) - \mbox{\boldmath
$\chi$}(\cdot,t,\bx)||_{0,[0,T]} + |||\bv_i -\bv|||_{0,\bar
\Lambda(T)} \to 0 \quad \mbox{ as } i \to \infty.
\end{eqnarray*}
We use (\ref{L7.4-2-1}) to show
\begin{equation}
[\mbox{\boldmath $\chi$}_i(\cdot,t,\bx) - \mbox{\boldmath
$\chi$}(\cdot,t,\bx)]_{0,\gamma,[0,T]}  \to 0, \quad \mbox{ as } i
\to \infty. \label{L7.4-2-2}
\end{equation}
By direct calculation we have
\begin{eqnarray*}
&& \frac{|(\mbox{\boldmath $\chi$}_i - \mbox{\boldmath
$\chi$})(s_1,t,\bx) - (\mbox{\boldmath $\chi$}_i - \mbox{\boldmath
$\chi$})(s_2,t,\bx)|}{|s_1 -s_2|^{\gamma}} \cr &&\leq ||\partial_s
\mbox{\boldmath $\chi$}_i(\cdot ,t,\bx) - \partial_s
\mbox{\boldmath $\chi$}(\cdot,t,\bx)||_{0,[0,T]}|s_1
-s_2|^{1-\gamma} \cr && \leq ||\partial_s \mbox{\boldmath
$\chi$}_i(\cdot ,t,\bx) - \partial_s \mbox{\boldmath
$\chi$}(\cdot,t,\bx)||_{0,[0,T]} T^{1-\gamma} \to 0 \quad \mbox{
as } i \to \infty.
\end{eqnarray*}
Next we show
\begin{equation}
[\partial_s \mbox{\boldmath $\chi$}_i(\cdot,t,\bx) - \partial_s
\mbox{\boldmath $\chi$}(\cdot,t,\bx)]_{0,\gamma,[0,T]} \to 0,
\quad \mbox{ as } i \to \infty. \label{L7.4-2-3}
\end{equation}
It follows from (\ref{char}) and (\ref{H1}) that
\begin{eqnarray*}
&& [\partial_s \mbox{\boldmath $\chi$}_i(\cdot ,t,\bx) -
\partial_s \mbox{\boldmath $\chi$}(\cdot,t,\bx)]_{0, \gamma,
[0,T]} \cr && \leq [ \bv_i(\mbox{\boldmath
$\chi$}_i(\cdot,t,\bx),\cdot) - \bv_i(\mbox{\boldmath
$\chi$}(\cdot,t,\bx),\cdot)]_{0,\gamma, [0,T]} + [
\bv_i(\mbox{\boldmath $\chi$}(\cdot,t,\bx),\cdot) -
\bv(\mbox{\boldmath $\chi$}(\cdot,t,\bx),\cdot)]_{0,\gamma, [0,T]}
\cr && \leq |||\nabla \bv_i|||_{0, \gamma, \bar \Lambda(T)} ||
\mbox{\boldmath $\chi$}_i(\cdot,t,\bx) - \mbox{\boldmath
$\chi$}(\cdot,t,\bx)||_{0,\gamma, [0,T]} + |||\bv_i
-\bv|||_{0,\gamma, \bar \Lambda(T)} \to 0 \quad \mbox{ as } i \to
\infty.
\end{eqnarray*}
Here we used (\ref{L7.4-2}) and (\ref{L7.4-2-2}). \newline

Finally we combine the estimates (\ref{L7.4-2}) - (\ref{L7.4-2-3})
to get
\[ |||\mbox{\boldmath $\chi$}_i(s,t,\bx) - \mbox{\boldmath $\chi$}(s,t,\bx)|||_{1,\gamma,[0,T]} \to 0 \quad \mbox{ as } i \to \infty. \]
\newline
\noindent (iv) By the triangle inequality, we have
\[
\frac{|\mbox{\boldmath $\alpha$}(\bx,t) - \mbox{\boldmath
$\alpha$}(\by,s)|}{|(\bx,t) - (\by,s)|} \leq
\frac{|\mbox{\boldmath $\alpha$}(\bx,t) - \mbox{\boldmath
$\alpha$}(\by,t)|}{| \bx - \by|} + \frac{|\mbox{\boldmath
$\alpha$}(\by,t) - \mbox{\boldmath $\alpha$}(\by,s)|}{| t - s|}.
\]
Here we used (\ref{L7.4-2}) and hypothesis (2) of this lemma. Next
observe that
\begin{eqnarray*}
&& \displaystyle  |\mbox{\boldmath $\alpha$}(\bx,t) -
\mbox{\boldmath $\alpha$}(\by,t)|  = |\mbox{\boldmath $\chi$}(0,t,
\bx) - \mbox{\boldmath $\chi$}(0,t, \by)| \cr && \displaystyle =
\Big| \int_0^1 \partial_{\xi}\mbox{\boldmath $\chi$}(0,t, \bx +
\xi(\by-\bx))d\xi \Big| = \Big| \int_0^1 \nabla_{\bx}
\mbox{\boldmath $\chi$}(0,t, \bx + \xi(\by-\bx)) \cdot (\by -\bx)
d\xi \Big| \cr && \displaystyle \leq \int_0^1 |\nabla_{\bx}
\mbox{\boldmath $\chi$}(0,t, \bx + \xi(\by-\bx))| |\by -\bx| d\xi
\leq 2|\by -\bx|.
\end{eqnarray*}
Here we used Lemma A1 (1):
\[ |||\nabla_{\bx} \mbox{\boldmath $\chi$}|||_{0,[0,T]\times [0,T] \times \bbr^2} \leq 2. \]
Similarly, we have
\[ |||\mbox{\boldmath $\alpha$}(\bx,t) - \mbox{\boldmath $\alpha$}(\by,t)|||_{0,[0,T]\times \bbr^2} \leq 2|t - s|. \]
Hence we have
\[ |\mbox{\boldmath $\alpha$}(\bx,t) - \mbox{\boldmath $\alpha$}(\by,s)| \leq 4|(\bx,t) - (\by,s)|. \]
\end{proof}

%
%
\begin{lemma}
Suppose $f$ is a scalar valued function defined on
$\Lambda_s^1(T;\bv)$ satisfying
\[ \sup_{0 \leq t \leq T} ||f(\cdot, t)||_{0,\gamma, \bar \Omega_s^1(t;\bv)} < \infty. \]
Then we have
\[
\displaystyle \Big[ \Big[ \int_0^t f(\mbox{\boldmath
$\chi$}(\xi,0, \mbox{\boldmath $\alpha$}(\bx,t)), \xi)d\xi
\Big]\Big]_{0,\gamma, \bar \Lambda_s^1(T;\bv)} \leq C_1(T) \Big(
\sup_{0 \leq t \leq T} ||f(\cdot, t)||_{0,\gamma, \bar
\Omega_s^1(t;\bv)} \Big ), \] where $[[~\cdot~]]_{0,\gamma, \bar
\Lambda_s^1(T;\bv)}$ is the H\"older seminorm  on the space-time
region, and
\[ C_1(T) := \Big(T^{1-\gamma} + 16^{\gamma}T \Big) = {\mathcal O}(T^{1-\gamma}). \]
If $f$ is in $C^{0,\gamma}(\bar \Lambda_s^1(T;\bv))$, then the
term $\sup_{0 \leq t \leq T} ||f(\cdot, t)||_{0,\gamma, \bar
\Omega_s^1(t;\bv)}$ can be replaced by $|||f|||_{0,\gamma, \bar
\Lambda_s^1(T;\bv)}$, i.e.,
\[
\displaystyle \Big[\Big[ \int_0^t f(\mbox{\boldmath $\chi$}(\xi,0,
\mbox{\boldmath $\alpha$}(\bx,t)), \xi)d\xi \Big]\Big]_{0,\gamma,
\bar \Lambda_s^1(T;\bv)} \leq C_1(T)|||f|||_{0,\gamma, \bar
\Lambda_s^1(T;\bv)}. \]
\end{lemma}
\begin{proof}
Let $(\bx,t)$ and $(\by,s)$ be two points in $\Lambda_s^1(T;\bv)$.
Without loss of generality, we assume that $s \leq t$.
\begin{eqnarray}
&& \frac{\Big|\int_0^t f(\mbox{\boldmath $\chi$}(\xi,0,
\mbox{\boldmath $\alpha$}(\bx,t)), \xi)d\xi - \int_0^s
f(\mbox{\boldmath $\chi$}(\xi,0, \mbox{\boldmath
$\alpha$}(\by,s)), \xi) d\xi \Big|}{|(\bx,t) - (\by,s)|^{\gamma}} \nonumber \\
&& \qquad \quad  \leq \frac{ \Big| \int_s^t f(\mbox{\boldmath
$\chi$}(\xi,0, \mbox{\boldmath
$\alpha$}(\bx,t)), \xi)d\xi \Big|}{|(\bx,t) - (\by,s)|^{\gamma}}  \nonumber \\
&& \qquad \quad + \frac{\int_0^s \Big | f(\mbox{\boldmath
$\chi$}(\xi,0, \mbox{\boldmath $\alpha$}(\bx,t)), \xi) -
f(\mbox{\boldmath $\chi$}(\xi,0, \mbox{\boldmath
$\alpha$}(\by,s)), \xi) \Big | d\xi}{|(\bx,t) -
(\by,s)|^{\gamma}}. \label{L7.5-1}
\end{eqnarray}
The terms on the right hand side of (\ref{L7.5-1}) can be treated
as follows:
\begin{eqnarray*}
&&\bullet~~\frac{ \Big| \int_s^t f(\mbox{\boldmath $\chi$}(\xi,0,
\mbox{\boldmath $\alpha$}(\bx,t)), \xi)d\xi \Big|}{|(\bx,t) -
(\by,s)|^{\gamma}} \leq (t-s)^{1-\gamma} |||f|||_{0,\bar
\Lambda_s(T;\bv)} \leq T^{1-\gamma} |||f|||_{0, \bar
\Lambda_s^1(T;\bv)}, \cr &&\bullet~~\Big | f(\mbox{\boldmath
$\chi$}(\xi,0, \mbox{\boldmath $\alpha$}(\bx,t)), \xi) -
f(\mbox{\boldmath $\chi$}(\xi,0, \mbox{\boldmath
$\alpha$}(\by,s)), \xi) \Big | \cr && \leq \Big( \sup_{0 \leq \xi
\leq T}[f(\cdot,\xi)]_{0,\gamma,\bar \Omega_s(\xi)} \Big)
|\mbox{\boldmath $\chi$}(\xi,0, \mbox{\boldmath $\alpha$}(\bx,t))
-  \mbox{\boldmath $\chi$}(\xi,0, \mbox{\boldmath
$\alpha$}(\by,s))|^{\gamma} \cr && \leq  \Big( \sup_{0 \leq \xi
\leq T}[f(\cdot,\xi)]_{0,\gamma,\bar \Omega_s(\xi)}
\Big)2^{\gamma} |\mbox{\boldmath $\alpha$}(\bx,t) -
\mbox{\boldmath $\alpha$}(\by,s) |^{\gamma} \cr && \leq
\Big(\sup_{0 \leq \xi \leq T}[f(\cdot,\xi)]_{0,\gamma,\bar
\Omega_s(\xi)} \Big) 8^{\gamma} |(\bx,t) - (\by,s)|^{\gamma}.
\end{eqnarray*}
Note that
\[ \max\Big \{ |||f|||_{0,\bar \Lambda_s^1(T;\bv)},~~
 \Big(\sup_{0 \leq t \leq T}[f(\cdot,t)]_{0,\gamma,\bar \Omega_s^1(t;\bv)} \Big) \Big \} \leq \sup_{0 \leq t \leq T}
 ||f(\cdot,t)||_{0,\gamma,\bar \Omega_s^1(t;\bv)}.\]
Hence we have the desired result. Furthermore if $f$ is in
$C^{0,\gamma}(\bar \Lambda_s^1(T;\bv))$, then the term $\sup_{0
\leq t \leq T} ||f(\cdot, t)||_{0,\gamma, \bar \Omega_s^1(t;\bv)}$
can be replaced by $|||f|||_{0,\gamma, \bar \Lambda_s^1(T;\bv)}$,
i.e.,
\[
\displaystyle \Big[\Big[ \int_0^t f(\mbox{\boldmath $\chi$}(\xi,0,
\mbox{\boldmath $\alpha$}(\bx,t)), \xi)d\xi \Big]\Big]_{0,\gamma,
\bar \Lambda_s^1(T;\bv)} \leq C_1(T)|||f|||_{0,\gamma, \bar
\Lambda_s^1(T;\bv)}. \]
\end{proof}
\vspace{0.5cm}
%
%
\begin{lemma}
Let $n$ be the solution of (\ref{ODE0}) given by (\ref{stepA}).
Then there exists a positive constant $T_1$ such that
 $n$ satisfies the a priori estimate:
\[ |||n|||_{0,\gamma, \bar \Lambda_s^1(T;\bv)} + \max_{|\alpha| = 1} |||\partial^{\alpha} n|||_{0,\gamma, \bar \Lambda_s^1(T;\bv)}
 + |||\partial_t n|||_{0,\gamma, \bar \Lambda_s^1(T;\bv)}
\leq R_1, \quad 0 < T \leq T_1, \] where $R_1$ is a positive
constant depending on $K_0, \delta^*$ and $\gamma$.
\end{lemma}
\begin{proof}
(i) Recall that $n$ satisfies
\begin{equation}
n(\bx,t) = n_0(\mbox{\boldmath $\alpha$}(\bx,t))
\exp\Big(-\int_0^t (\nabla \cdot \bv)(\mbox{\boldmath
$\chi$}(\xi,0, \mbox{\boldmath $\alpha$}(\bx,t)), \xi)d\xi \Big),
\quad \mbox{ for } (\bx,t) \in \Lambda_s^1(T;\bv). \label{L7.6-1}
\end{equation}
Since $\bv \in {\mathcal B}_{T}$, we have
\[ |||\nabla \cdot \bv|||_{0,\gamma,\bar \Lambda(T)} \leq 6K_0 \delta^*  ~~\mbox{ in } \Lambda(T) \]
and hence (\ref{L7.6-1}) implies
\[
\displaystyle |||n|||_{0,\bar \Lambda_s^1(T;\bv)} \leq e^{6TK_0
\delta^*} ||n_0||_{0,\bar \Omega_s^1(0;\bv)}.
\]
Furthermore if we assume  $T_1$ is sufficiently small enough to
satisfy
\begin{equation}
e^{6T_1K_0 \delta^*} \leq 2 \label{L7.6-2}
\end{equation}
then we have
\begin{equation}
\displaystyle |||n|||_{0,\bar \Lambda_s^1(T;\bv)} \leq 2
||n_0||_{0,\bar \Omega_s^1(0;\bv)}. \label{L7.6-3}
\end{equation}
Next we show that $n_0(\mbox{\boldmath $\alpha$}(\bx,t))$ is in
$C^{0,\gamma}(\bar \Lambda_s^1(T;\bv))$. Let $(\bx,t)$ and
$(\by,s)$ be points in $\Lambda_s^1(T;\bv)$.
 Without loss of generality, we assume $s \leq t$. Then we have
\begin{eqnarray*}
  \displaystyle \frac{|n_0(\mbox{\boldmath $\alpha$}(\bx,t)) - n_0(\mbox{\boldmath $\alpha$}(\by,s))|}{|(\bx,t) - (\by,s)|^{\gamma}}
  &=& \frac{|n_0(\mbox{\boldmath $\alpha$}(\bx,t)) - n_0(\mbox{\boldmath $\alpha$}(\by,s))|}
{|\mbox{\boldmath $\alpha$}(\bx,t) - \mbox{\boldmath
$\alpha$}(\by,s)|^{\gamma}}
 \Big( \frac{|\mbox{\boldmath $\alpha$}(\bx,t) - \mbox{\boldmath $\alpha$}(\by,s)|}{|(\bx,t) - (\by,s)|} \Big)^{\gamma} \cr
 \displaystyle &\leq& [n_0]_{0,\gamma, \bar \Omega^1_s(0;\bv)} 4^{\gamma},
\end{eqnarray*}
and hence
\begin{equation}
[[n_0(\mbox{\boldmath $\alpha$})]]_{0,\gamma, \bar
\Lambda_s^1(T;\bv)} \leq [n_0]_{0,\gamma, \bar \Omega_s^1(0;\bv)}
4^{\gamma}. \label{L7.6-4}
\end{equation}
On the other hand, it follows from Lemma A.2 and the fact that
$\bv \in {\mathcal B}(T)$ (see (${\mathcal D}(2)$) in Definition
7.1) that
\[ \Big[\Big[ -\int_0^t (\nabla \cdot \bv)(\mbox{\boldmath $\chi$}(\xi,0, \mbox{\boldmath
$\alpha$}(\bx,t)), \xi)d\xi \Big]\Big ]_{0,\gamma,\bar
\Lambda_s^1(T;\bv))} \leq 6C_1(T)K_0 \delta^*. \] We use
(\ref{H2}) and (\ref{L7.6-2}) to get
\begin{equation}
\Big[ \Big[ \exp\Big(-\int_0^t (\nabla \cdot \bv)(\mbox{\boldmath
$\chi$}(\xi,0, \mbox{\boldmath $\alpha$}(\bx,t)), \xi)d\xi \Big)
\Big]\Big ]_{0,\gamma,\bar \Lambda_s^1(T;\bv))} \leq 12C_1(T)K_0
\delta^*, \label{L7.6-5}
\end{equation}
and then use (\ref{H1}),  (\ref{L7.6-2}), (\ref{L7.6-4}) and
(\ref{L7.6-5}) to find
\begin{equation}
[[n]]_{0,\gamma, \bar \Lambda_s^1(T)} \leq 2[n_0]_{0,\gamma, \bar
\Omega_s^1(0;\bv)} 4^{\gamma} + 12 ||n_0||_{0,\bar
\Omega_s^1(0;\bv)} C_1(T)K_0 \delta^*. \label{L7.6-5-1}
\end{equation}
Since $C_1(T_1) = {\mathcal O}(T_1^{1-\gamma})$,  we have for
$T_1$ sufficiently small that
\begin{equation}
12 C_1(T)K_0 \delta^* \leq 1, \quad T \leq T_1, \label{L7.6-6}
\end{equation}
so that (\ref{L7.6-5-1}) implies
\begin{equation}
[[n]]_{0,\gamma, \Lambda_s^1(T;\bv)} \leq 2[n_0]_{0,\gamma, \bar
\Omega_s^1(0;\bv)} 4^{\gamma} + ||n_0||_{0,\bar \Omega_s^1(0)}.
\label{L7.6-7}
\end{equation}
Finally combine (\ref{L7.6-3}) and (\ref{L7.6-7}) to get the
desired bound
\begin{eqnarray}
|||n|||_{0,\gamma, \bar \Lambda_s^1(T)} &\leq& \max\{ 2^{2\gamma + 1}, 3 \} ||n_0||_{0,\gamma, \bar \Omega_s^1(0)}, \quad 0 < T \leq T_1,  \nonumber \\
                                      &\leq& \max\{ 2^{2\gamma +1}, 3 \} \delta^*. \label{L7.6-8}
\end{eqnarray}

\vspace{0.5cm}

(ii) We now need to estimate space derivatives of $n$.
Differentiate the continuity equation
\[ \displaystyle \partial_t n + \sum_{i=1}^{2} \partial_{x_i}(n v_i) = 0 \]
with respect to $x_j$ to find
\begin{equation}
\displaystyle \frac{D (\partial_{x_j} n)}{Dt} = -\Big(
\sum_{i=1}^{2} \partial_{x_ix_j}^2 v_i \Big) n -
\Big(\sum_{i=1}^{2} \partial_{x_i} v_i \Big) \partial_{x_j} n -
\Big(\sum_{i=1}^{2} \partial_{x_i} n  \partial_{x_j} v_i \Big),
~~j=1,2. \label{L7.6-9}
\end{equation}
Here $\frac{D}{Dt} = \partial_t + \bv \cdot \nabla_{x}$. \newline

Integrate (\ref{L7.6-9}) along the characteristic curve
$\mbox{\boldmath $\chi$}$ to obtain
\begin{eqnarray}
\partial_{x_j} n(\bx,t) &=& \partial_{x_j} n_0(\mbox{\boldmath $\alpha$}(\bx,t)) \nonumber \\
 &-& \sum_{i=1}^2 \int_0^t  \Big( n \partial_{x_ix_j}^2 v_i   + \partial_{x_i} v_i \partial_{x_j} n
 + \partial_{x_i}n \partial_{x_j} v_i \Big)(\mbox{\boldmath $\chi$}(\xi,0, \mbox{\boldmath $\alpha$}(\bx,t)), \xi)d\xi.
\label{L7.6-11}
\end{eqnarray}
Since $\bv$ satisfies
\[
\displaystyle \max_{i=1,2} \Big( \max_{|\alpha|=1}
|||\partial^{\alpha} v_i|||_{0,\gamma, \bar \Lambda^1(T;\bv)}  +
 \max_{|\alpha|=2}  |||\partial^{\alpha} v_i|||_{0,\gamma, \bar \Lambda(T)} \Big)
\leq 3K_0\delta^*, \quad \mbox{ by the definition of } {\mathcal
B}(T) \] and $n$ satisfies
\[ \displaystyle |||n|||_{0,\bar \Lambda_s^1(T;\bv)} \leq 2 ||n_0||_{0,\bar \Omega_s^1(0;\bv)}, \]
for T sufficiently small by (\ref{L7.6-3}), we have from
(\ref{L7.6-11}) that
\begin{eqnarray}
&& \max_{|\alpha|=1} |||\partial^{\alpha} n|||_{0, \bar \Lambda_s^1(T;\bv)} \nonumber  \\
&& \qquad \leq \max_{|\alpha|=1}||\partial^{\alpha} n_0||_{0, \bar
\Omega_s^1(0;\bv)} + 6TK_0 \delta^* ||n_0||_{0,\bar
\Omega_s^1(0;\bv)}
                                    + 12K_0 \delta^* T \max_{|\alpha|=1}|||\partial^{\alpha} n|||_{0, \bar \Lambda_s^1(T;\bv)}. \nonumber \\
\label{L7.6-11-1}
\end{eqnarray}
We assume $T_1$ is sufficiently small so that
\begin{equation}
 K_0\delta^* T_1 \leq \frac{1}{24}, \label{L7.6-12}
\end{equation}
then we have from (\ref{L7.6-12}) that
\begin{equation}
\max_{|\alpha|=1}|||\partial^{\alpha} n|||_{0, \bar
\Lambda_s^1(T;\bv)}
 \leq 2\max_{|\alpha|=1}||\partial^{\alpha} n_0||_{0,\bar \Omega_s^1(0;\bv)} + \frac{1}{2} ||n_0||_{0,\bar \Omega_s^1(0;\bv)} \quad \mbox{ for }
T \leq T_1. \label{L7.6-13}
\end{equation}
Next we estimate $[[\partial^{\alpha} n]]_{0,\gamma, \bar
\Lambda_s^1(T;\bv)}, |\alpha|=1$ using Lemma A.2 and (\ref{H1}).
\newline

By direct calculation, we have following estimates: For $T >0$
sufficiently small, we have
\begin{eqnarray}
&& \bullet~~[[\partial_{x_j} n_0(\mbox{\boldmath
$\alpha$}(\bx,t))]]_{0,\gamma, \bar \Lambda_s^1(T;\bv)} \leq
4^{\gamma}
\max_{|\alpha|=1} [\partial^{\alpha} n_0]_{0,\gamma, \bar \Omega_s^1(0;\bv)}, \label{L7.6-13-1} \\
&& \bullet~~[[\partial_{x_ix_j}^2 v_i  n]]_{0,\gamma, \bar \Lambda_s^1(T;\bv)} \leq 6K_0 (\delta^*)^2 \max\{2^{2\gamma + 1}, 3\}, \label{L7.6-13-2} \\
&& \bullet~~ [[\partial_{x_i} v_i \partial_{x_j} n]]_{0,\gamma,
\bar \Lambda_s^1(T;\bv)} \leq 3K_0 \delta^*[[\nabla n]]_{0,\gamma,
\bar \Lambda_s^1(T;\bv)} +
3K_0 \delta^* |||\nabla n |||_{0,\bar \Lambda_s^1(T;\bv)}, \label{L7.6-13-3} \\
&& \bullet~~[[ \partial_{x_i}n \partial_{x_j} v_i ]]_{0,\gamma,
\bar \Lambda_s^1(T;\bv)} \leq 3K_0 \delta^* \Big( [[\nabla
n]]_{0,\gamma, \bar \Lambda_s^1(T;\bv)}
 + \max_{|\alpha|=1} |||\partial^{\alpha} n|||_{0,\bar \Lambda_s^1(T;\bv)} \Big). \label{L7.6-13-4}
\end{eqnarray}
We combine estimates (\ref{L7.6-13-1}) - (\ref{L7.6-13-4})  to get
\begin{eqnarray}
\max_{|\alpha|=1} [[\partial^{\alpha} n]]_{0,\gamma, \bar
\Lambda_s^1(T;\bv)} &\leq& 4^{\gamma} \max_{|\alpha|=1}
[\partial^{\alpha} n_0]_{0,\gamma, \bar \Omega_s^1(0;\bv)}
                   + 12K_0 \delta^* C_1(T) \nonumber \\
                    &\times& \Big(\delta^* \max\{2^{2\gamma +1},3\} + \max_{|\alpha|=1} [[\partial^{\alpha} n]]_{0,\gamma, \bar \Lambda_s^1(T;\bv)}
                    + \max_{|\alpha|=1} |||\partial^{\alpha} n|||_{0,\bar \Lambda_s^1(T;\bv)}\Big). \label{L7.6-14}
\end{eqnarray}
We assume $T_1$ is sufficiently small so that
\begin{equation}
K_0 \delta^* C_1(T_1) \leq \frac{1}{24} \label{L7.6-15}
\end{equation}
and hence for $T \in (0, T_1]$, (\ref{L7.6-14}) implies
\begin{eqnarray}
\max_{|\alpha|=1}[[\partial^{\alpha} n]]_{0,\gamma, \bar
\Lambda_s^1(T;\bv)} &\leq& 2^{2\gamma + 1} \max_{|\alpha|=1}
[\partial^{\alpha} n_0]_{0,\gamma, \bar \Omega_s^1(0;\bv)} + \delta^* \max\{24^{\gamma}, 3\} \nonumber \\
 &+& 2 \max_{|\alpha|=1} ||\partial^{\alpha} n_0||_{0,\bar \Omega_s^1(0;\bv)} + \frac{1}{2} ||n_0||_{0,\bar \Omega_s^1(0;\bv)}. \label{L7.6-16}
\end{eqnarray}
We combine (\ref{L7.6-13}) and (\ref{L7.6-16}) to obtain
\begin{equation}
\max_{|\alpha|=1} |||\partial^{\alpha} n|||_{0,\gamma, \bar
\Lambda_s^1(T;\bv)} \leq 2 \delta^* \max \{2^{2\gamma + 1}, 4 \}.
\label{L7.6-17}
\end{equation}
(iii) Now we estimate the time derivative of $n$.  Recall that $n$
satisfies
\begin{equation}
\partial_t n + \nabla \cdot (n \bv) = 0. \label{L7.6-18}
\end{equation}
Next we use (\ref{L7.6-8}), (\ref{L7.6-17}) and (\ref{L7.6-18}) to
see
\begin{eqnarray}
||\partial_t n||_{0, \gamma, \bar \Lambda_s^1(T;\bv)} &\leq& 2
\max_{|\alpha|=1} ||\partial^{\alpha} n||_{0,\gamma,\bar
\Lambda_s^1(T;\bv)}  \cdot |||\bv|||_{0,\gamma,\bar \Lambda(T)} +
||n||_{0,\gamma,\bar \Lambda_s^1(T;\bv))}
 || \nabla \cdot \bv||_{0,\gamma, \bar \Lambda(T))} \nonumber \\
  &\leq& 15K_0 (\delta^*)^2 \max\{2^{2\gamma + 1}, 4\}. \label{L7.6-19}
\end{eqnarray}
Finally we set
\[
 R_1(K_0, \delta^*, \gamma) := \max\{2^{2\gamma + 1}, 3 \} \delta^*  + 2\delta^* \max\{ 2^{2\gamma + 1}, 4\}
  + 15K_0 (\delta^*)^2 \max\{2^{2\gamma + 1}, 4\}
\]
to see that (\ref{L7.6-8}), (\ref{L7.6-17}) and (\ref{L7.6-19})
imply the desired result.
\end{proof}
\end{subsubsection}

\begin{subsubsection}{{\bf A priori estimates for {\it Step 2}}} In this part, we will give existence, uniqueness and regularity for the function $\zeta$
as given in {\it Step 2} of Section 7.2. \newline

Recall from {\it Step 2}, $\zeta$ satisfies the exterior Neumann
problem for Laplace's equation at given time $t \in [0,T]$:
\begin{equation}
\begin{cases}
\Delta \zeta(\cdot,t)  = 0, \quad \bx \in \Omega_1, \\
\displaystyle \nabla \zeta \cdot \mbox{\boldmath$\nu$}_0 = h_0,
\quad \bx \in \partial \Omega_0 \quad \mbox{ and } \quad \lim_{|x|
\to \infty} \nabla \zeta = {\bf 0}.
\end{cases} \label{HH1}
\end{equation}
%
%
\begin{lemma}
$h_0 \in C^{1,\gamma}(\partial \Omega_0 \times [0, T])$ and
satisfies
\[ |||h_0|||_{1,\gamma,\partial \Omega_0 \times [0, T]} \leq \delta^* + 6\bar C_0 K_0 R_1 (\delta^*)^2, \]
where $\bar C_0$ is a positive constant.
\end{lemma}
\begin{proof}
Recall that $h_0 = \partial_t g - (n \bv) \cdot \mbox{\boldmath
$\nu$}_0$. Then
\begin{eqnarray}
\displaystyle |||h_0|||_{1,\gamma,\partial \Omega_0 \times [0,
T]}&\leq& |||\partial_t g|||_{1,\gamma,\partial \Omega_0 \times
[0, \infty)} \cr &+& |||nv_1 \nu_{01} |||_{1,\gamma,\partial
\Omega_0 \times [0, T]} + |||nv_2 \nu_{02} |||_{1,\gamma,\partial
\Omega_0 \times [0, T]}.  \label{Ha}
\end{eqnarray}
Since the product of H\"older continuous functions is again
H\"older continuous (see \cite{G-T}, pg. 53), we have
\begin{eqnarray}
&& |||nv_1 \nu_{01} |||_{1,\gamma,\partial \Omega_0 \times [0, T]} + |||nv_2 \nu_{02} |||_{1,\gamma,\partial \Omega_0 \times [0, T]} \nonumber \\
&& \qquad \quad  \leq \bar C_{0} \Big( |||n|||_{1,\gamma,\partial
\Omega_0 \times [0, T]} |||v_1|||_{1,\gamma,\partial \Omega_0
\times [0, T]} |||\nu_{01}|||_{1,\gamma,\partial \Omega_0 \times
[0, T]}
 \nonumber \\
&& \qquad \quad + |||n|||_{1,\gamma,\partial \Omega_0 \times [0,
T]} |||v_2|||_{1,\gamma,\partial \Omega_0 \times [0, T]}
|||\nu_{02}|||_{1,\gamma,\partial \Omega_0
\times [0, T]} \Big) \nonumber \\
&& \qquad \quad \leq 6\bar C_0 K_0 R_1 (\delta^*)^2. \label{HH2}
\end{eqnarray}
Here $\bar C_0$ is a positive constant and we used Lemma A.3, $
\displaystyle \max_{i=1,2}|||v_i|||_{1,\gamma, \bar \Lambda(T)}
\leq 3K_0 \delta^*$,  and inequalities
\[ |||n|||_{1,\gamma,\partial \Omega_0 \times [0, T]} \leq |||n|||_{1,\gamma,\bar \Lambda_s^1(T;\bv)} \quad \mbox{ and }
\quad |||v_i|||_{1,\gamma,\partial \Omega_0 \times [0, T]} \leq
|||v_i|||_{1,\gamma,\bar \Lambda(T)}. \] Hence in (\ref{Ha}), we
use (\ref{HH2}) and the assumption (A2) of Section 7.2 to get
\begin{equation}
|||h_0|||_{1,\gamma,\partial \Omega_0 \times [0, T]} \leq \delta^*
+ 6\bar C_0 K_0 R_1 (\delta^*)^2. \label{HH3}
\end{equation}
\end{proof}

Recall the annulus region (\ref{Set1}) in Section 7.2:
\[ \Omega_* = \{ \bx \in \bbr^2: \frac{\delta_{*2}}{2} < |\bx| < 2\delta^* \}. \]

The following existence and uniqueness result of two-dimensional
exterior Neumann problem (\ref{HH1}) is due to the results of Bers
\cite{Bers} and Finn-Gilbarg \cite{F-G}.

%
%
\begin{lemma}
Suppose the boundary data $h_0$ is in $C^{1,\gamma}(\partial
\Omega_0 \times [0, \infty))$ as provided by Lemma A.4.
 Then there exists a unique solution $\zeta$ up to constant of (\ref{HH1}) satisfying the following
estimates: For the compactly supported subset $\Omega_*$ of
$\Omega_1$, we have
\begin{enumerate}
\item
\[ \displaystyle
|||\nabla \zeta|||_{1,\gamma, \bar \Omega_* \times [0, T]} +
\sup_{0 \leq t \leq T} \max_{|\alpha|=3} |||\partial^{\alpha}
\zeta(\cdot,t)|||_{0,\gamma, \bar \Omega_*} \leq \bar R_1, \quad 0
\leq t \leq T,
\]
where $\bar R_1$ is a positive constant which depends only  on
$\Omega_0$, $\delta^*$. \item Let $h_0^{(n)} \in
C^{1,\tau}(\partial \Omega_0 \times [0, \infty))$ be a sequence of
boundary data  satisfying the bound (\ref{HH3}) and
\[ h_0^{(n)} \to h_0 \quad \mbox{ in } C^{1,\tau}(\partial \Omega_0 \times [0, \infty)) \quad \mbox{ as } n \to \infty. \]
Suppose $\nabla \zeta^{(n)}$ and $\nabla \zeta$ are the
corresponding solutions to the above exterior Neumann problem
(\ref{HH1}) for data $h_0^{(n)}, h_0$ respectively. Then we have
\[ \nabla \zeta^{(n)} \to \nabla \zeta \quad \mbox{ in } (C^{1,\tau}(\bar \Omega_* \times [0,T]))^2 \quad \mbox{ as } n \to \infty. \]
\end{enumerate}
\end{lemma}
\begin{proof}
(i) Since $\Omega_*$ is compactly supported in $\Omega_1$,  it
follows from the interior Schauder estimates (\cite{G-T}, Section
6.1), we have
\begin{eqnarray}
&& \displaystyle ||\nabla \zeta(\cdot,t)||_{1,\gamma, \bar
\Omega_*}
+ \max_{|\alpha|=3} ||\partial^{\alpha} \zeta(\cdot,t)||_{0, \gamma, \bar \Omega_*} \nonumber \\
&& \hspace{2cm} \leq C_0(\Omega_0, \Omega_*)|||h_0|||_{1,\gamma,\partial \Omega_0 \times [0, T]} \nonumber \\
&& \hspace{2cm} \leq C_0(\Omega_0, \Omega_*) (\delta^* + 6\bar C_0
K_0 R_1 (\delta^*)^2), \quad 0\leq t \leq T. \label{LA.4-1}
\end{eqnarray}
Let $0 \leq s < t$. Then it follows from Laplace's equation and
the boundary condition that
\[
\begin{cases}
 \Delta \Big( \frac{\zeta(\bx,t) - \zeta(\bx,s)}{|t-s|^{\gamma}}  \Big) = 0, \quad \bx \in \Omega_1,\quad 0\leq s < t \leq T, \cr
 \nabla \Big(\frac{\zeta(\bx,t) - \zeta(\bx,s)}{|t-s|^{\gamma}} \Big) \cdot \mbox{\boldmath$\nu$}_0  = \frac{h_0(x,t) - h_0(\bx,s)}{|t-s|^{\gamma}}, \quad
\bx \in \partial \Omega_0.
\end{cases}
\]
We now apply the global Schauder estimate (\cite{G-T}, Section
6.2) to get
\begin{eqnarray}
&& \frac{|\nabla \zeta(\bx,t) - \nabla \zeta(\bx,s)|}{|t-s|^{\gamma}} \nonumber \\
&& \hspace{2cm} \leq C_1(\Omega_0, \Omega_*) |||h_0|||_{1,\gamma,\partial \Omega_0 \times [0, T]} \nonumber \\
&& \hspace{2cm} \leq C_1(\Omega_0, \Omega_*) (\delta^* + 6\bar C_0
K_0 R_1 (\delta^*)^2), \quad \bx \in \Omega_1. \label{LA.4-2}
\end{eqnarray}
We take the supremum over $ t \not = s$ to get
\begin{equation}
\displaystyle  \sup_{ t \not = s} \frac{|\nabla \zeta(\bx,t) -
\nabla \zeta(\bx,s)|}{|t-s|^{\gamma}} \leq C_1(\Omega_0, \Omega_*)
(\delta^* + 6\bar C_0 K_0 R_1 (\delta^*)^2), \qquad \bx \in
\Omega_1. \label{LA.4-3}
\end{equation}
Let $(\bx,t) \not = (\by,s)$ and  without loss of generality,
assume that $\bx \not = \by, t \not = s$. From (\ref{LA.4-1}) and
(\ref{LA.4-3}), the H\"older quotient satisfies
\begin{eqnarray*}
\frac{|\nabla \zeta(\bx,t) - \nabla \zeta(\by,s)|}{|(\bx,t)
-(\by,s)|^{\gamma}} &\leq& \frac{|\nabla \zeta(\bx,t) - \nabla
\zeta(\bx,s)|}{|t-s|^{\gamma}} + \frac{|\nabla \zeta(\bx,s) -
\nabla \zeta(\by,s)|}{|\bx -\by|^{\gamma}} \cr &\leq& [\nabla
\zeta(\bx, \cdot)]_{0,\gamma, [0,T]} + [\nabla
\zeta(\cdot,s)]_{0,\gamma,\Omega_s(s)} \cr &\leq& (C_0(\Omega_0,
\Omega_*) + C_1(\Omega_0, \Omega_*))(\delta^* + 6\bar C_0 K_0 R_1
(\delta^*)^2).
\end{eqnarray*}
Taking sup over the time-space region $\Omega_* \times [0,T]$, we
have
\begin{equation}
[\nabla \zeta ]_{0,\gamma,\bar \Omega_* \times [0, T]} \leq
(C_0(\Omega_0, \Omega_*) + C_1(\Omega_0, \Omega_*))(\delta^* +
6\bar C_0 K_0 R_1 (\delta^*)^2). \label{LA.4-3}
\end{equation}
Similarly we can estimate $\displaystyle \max_{|\alpha|=2}
|||\partial^{\alpha} \zeta|||_{0,\gamma, \bar \Omega_* \times
[0,T]}$ to get
\begin{equation}
\max_{|\alpha|=2} |||\partial^{\alpha} \zeta|||_{0,\gamma, \bar
\Omega_* \times [0,T]} \leq C_2(\Omega_0, \Omega_*) (\delta^* +
6\bar C_0 K_0 R_1 (\delta^*)^2). \label{LA.4-4}
\end{equation}
Finally we combine (\ref{LA.4-1}), (\ref{LA.4-3}) and
(\ref{LA.4-4}) to get
\[ \displaystyle
|||\nabla \zeta|||_{1,\gamma, \bar \Omega_* \times [0, T]} +
\sup_{0 \leq t \leq T} \max_{|\alpha|=3} |||\partial^{\alpha}
\zeta(\cdot,t)|||_{0,\gamma, \bar \Omega_*} \leq \bar R_1,
\]
where $\bar R_1 := (C_0(\Omega_0, \Omega_*) + C_1(\Omega_0,
\Omega_*) + C_2(\Omega_0, \Omega_*))(\delta^* + 6\bar C_0 K_0 R_1
(\delta^*)^2) $. \newline

(ii) The difference $\zeta^{(n)} - \zeta$ satisfies
\[
\begin{cases}
\Delta (\zeta^{(n)}(\cdot,t) - \zeta(\cdot,t))  = 0, \quad \bx \in \Omega_1, \\
\displaystyle \nabla (\zeta^{(n)} - \zeta) \cdot
\mbox{\boldmath$\nu$}_0 = h_0^{(n)} -  h_0, \quad \bx \in \partial
\Omega_0 \quad \mbox{ and } \quad \zeta^{(n)} - \zeta =0 \quad
\mbox{ on } \partial B(0, 3\delta^*).
\end{cases}
\]
By the Schauder estimates (Section 6.2 in \cite{G-T}), we have
\[ ||\nabla \zeta^{(n)}(\cdot,t) - \nabla \zeta(\cdot,t) ||_{1,\tau, \bar \Omega_1} \leq C||h_0^{(n)} - h_0||_{1,\tau,\partial \Omega_0}.\]
Letting $n \to \infty$, it follows from  the above inequality and
hypothesis (2) of this lemma that
\begin{equation}
\nabla \zeta^{(n)}(\cdot,t) \to \nabla \zeta(\cdot,t) \quad \mbox{
in } (C^{1,\tau}(\bar \Omega_1))^2, \quad 0 \leq t \leq T.
\label{LA.4-5}
\end{equation}
For the time-estimates we apply the same method as in (i) to get
\begin{equation}
\nabla \zeta^{(n)}(\bx,\cdot) \to \nabla \zeta(\bx,\cdot) \quad
\mbox{ in } (C^{1,\tau}([0,T]))^2, \quad \bx  \in \bar \Omega_*.
\label{LA.4-6}
\end{equation}
We combine (\ref{LA.4-5}) and (\ref{LA.4-6}) to see
\[ \nabla \zeta^{(n)} \to \nabla \zeta \quad \mbox{ in } (C^{1,\tau}(\bar \Omega_1 \times [0,T]))^2. \]
This yields
\[ \nabla \zeta^{(n)} \to \nabla \zeta \quad \mbox{ in } (C^{1,\tau}(\bar \Omega_* \times [0,T]))^2. \]
\end{proof}
\end{subsubsection}
%
%
\begin{subsubsection}{{\bf A priori estimates for {\it Step 3}}} We now need to give existence, uniqueness and regularity for the interface of {\it Step 3} of Section 7.2.
\begin{lemma}
1. Assume that  $\zeta$ satisfies estimate (1) of Lemma A.5:
\[ \displaystyle
||\nabla \zeta||_{1,\gamma, \bar \Omega_* \times [0, T]} +
\sup_{0\leq t \leq T} \max_{|\alpha|=3} ||\partial^{\alpha}
\zeta(\cdot,t)||_{0, \gamma, \bar \Omega_*} \leq \bar R_1. \] Then
there exists a unique solution for the interface system
(\ref{EQ7-2-8}) satisfying
\[
(\theta, n_s, r) \in (C^{1,\gamma}(\bbr \times [0, T]))^3 \quad
\mbox{ and } \quad (\partial_{\beta} \theta, \partial_{\beta}n_s,
\partial_{\beta}r) (\cdot,t) \in (C^{2,\gamma}(\bbr))^3, \quad t
\in [0, T]. \] Moreover, we have
\begin{eqnarray*}
&& \bullet~~||\theta||_{1,\gamma, \bbr \times [0, T]} +
||n_s||_{1,\gamma, \bbr \times [0, T]} + ||r||_{1,\gamma, \bbr
\times [0, T]} \cr && \hspace{3cm} \leq 2
\Big(||\theta_{0}||_{1,\gamma, \bbr} + ||n_{s0}||_{1,\gamma, \bbr}
+ ||r_{0}||_{1,\gamma, \bbr} \Big), \cr &&
\bullet~~||\partial_{\beta} \theta(\cdot,t)||_{2,\gamma, \bbr} +
||\partial_{\beta} n_s(\cdot,t)||_{2,\gamma, \bbr} +
||\partial_{\beta} r(\cdot,t)||_{2,\gamma, \bbr} \cr &&
\hspace{3cm} \leq 2 \Big(||\partial_{\beta}
\theta_{0}||_{2,\gamma, \bbr} + ||\partial_{\beta}
n_{s0}||_{2,\gamma, \bbr} + ||\partial_{\beta} r_{0}||_{2,\gamma,
\bbr} \Big).
\end{eqnarray*}
2. Let
\[ \nabla \zeta^{(n)} \to \nabla \zeta \quad \mbox{ in } (C^{1,\tau}(\bar \Omega_* \times [0, T]))^2 \quad \mbox{ as given by (2) of Lemma A.4}, \]
and let $(\theta^{(n)}, n_s^{(n)}, r^{(n)})$ and $(\theta, n_s,
r)$ be the solutions of the sheath system corresponding to $\nabla
\zeta^{(n)}$ and $\nabla \zeta$ respectively. Then we have
\[ (\theta^{(n)}, n_s^{(n)}, r^{(n)}) \to (\theta, n_s, r) \quad \mbox{ in } (C^{1,\tau}(\bbr \times [0,T]))^3 \quad \mbox{ as } n \to \infty. \]
\end{lemma}
\begin{proof}
The result is just continuity with respect to data for the
hyperbolic system (\ref{EQ7-2-8}). The proof of convergence in
$C^1(\bbr \times [0,T])$ follows from the argument in \cite{Do}.
The proof of H\"older norms $C^{1,\gamma}(\bbr \times [0,T])$ is
similar to that of \cite{Do} (see \cite{L-Y}.
\end{proof}
\end{subsubsection}

\begin{subsubsection}{{\bf A priori estimates for {\it Step 4}}} We next present the existence, uniqueness and regularity of the ion density $n$ in the region
$\Lambda_s^2(T)$ given by {\it Step 4} of Section 7.2.
%
%
\begin{lemma}
Let $n$ be the ion-density obtained from {\it Step 4}. Then the
formulas in {\it Step 4}, namely $n$ satisfies the differential
equations:
\[
\begin{cases}
\displaystyle   \frac{d}{ds} \mbox{\boldmath $\chi$}(s, t_0,
\mbox{\boldmath $\alpha$})
  = \bv (\mbox{\boldmath $\chi$}(s,t_0, \mbox{\boldmath $\alpha$}),s), \quad s > t_0, \cr
 \displaystyle \frac{d}{ds} \ln n(\mbox{\boldmath $\chi$}(s,t_0, \mbox{\boldmath $\alpha$}),s)
= - (\nabla \cdot \bv)(\mbox{\boldmath
$\chi$}(s,t_0,\mbox{\boldmath $\alpha$}),s),
\end{cases}
\]
subject to initial and boundary data:
\[ \mbox{\boldmath $\chi$}(t_0, t_0, \mbox{\boldmath
$\alpha$}) = \mbox{\boldmath $\alpha$} \quad \mbox{ and } \quad
n(\mbox{\boldmath $\alpha$}, t_0) =
\begin{cases}
n_0(\mbox{\boldmath $\alpha$)} \quad  t_0 = 0, \cr
n_s(\mbox{\boldmath $\alpha$}, t_0) \quad t_0 > 0,
\end{cases}
\]
are indeed valid. Furthermore for sufficiently small $T$, the
following estimates hold:
\[ n \in C^{1,\gamma}(\Lambda_s^2(T;\bv)) \quad \mbox{ and }\quad
 ||n||_{1,\gamma; \Lambda_s^2(T;\bv)} \leq R_2, \]
where $R_2$ is a positive constant depending only on $K_0,
\delta^*, \gamma$.
\end{lemma}
\begin{proof}
The proof follows from Remark A.1, i.e. since backward
characteristics starting at a point $(\bx,t) \in
\Lambda_s^2(T;\bv)$ can be traced back to a point
$(\mbox{\boldmath $\alpha$},0)$ in the absence of the sheath
interface, the presence of the sheath interface means backward
characteristics must hit either a point in $\Omega_s^2(0;\bv)$ or
a point in the sheath interface. Furthermore, the segment of
backwards characteristic between $(\bx,t)$ and  $(\mbox{\boldmath
$\alpha$},0)$  can hit the sheath interface at most once. Indeed,
the backwards characteristic can enter but not exit the domain
$\Lambda_s(T;\bv)=\Lambda_s^1(T;\bv)\cap \Lambda_s^2(T;\bv)$
through the interface surface at time $0<t<T$. This is because
initially $\bu_0 = -\mbox{\boldmath $\nu$}$ on ${\mathcal S}(0)$,
hence $|\bv\cdot \mbox{\boldmath $\nu$}+1|<\varepsilon$ on
${\mathcal S}(t)$ for  $0<t<T$ by (A1) and Theorem
\ref{existInterfTheorem}, where $T>0$ is sufficiently small
depending only on initial data, boundary data, and
$\varepsilon>0$. By choosing $\varepsilon>0$ sufficiently small,
the vector field for the characteristic $\displaystyle
{d\mbox{\boldmath $\chi$}\over dt}=\bv(\mbox{\boldmath $\chi$},
t)$ always points into the domain $\Lambda_s(T;\bv)$ at the point
of intersection with  the sheath interface.

Hence the formulas follow from (\ref{ODE0}).  Furthermore the
regularity estimates in the statement of the lemma can be obtained
in a similar manner as in Lemma A.3.
\end{proof}
We combine Lemma A.3 and Lemma A.7 to get the regularity result
for $n$ in the sheath region.
%
%
\begin{lemma}
For sufficiently small $T$, we have
\[ n \in C^{1,\gamma}(\bar \Lambda_s(T;\bv)) \quad \mbox{ and } \quad  ||n||_{1,\gamma, \bar  \Lambda_s(T;\bv)} \leq R_1 + R_2. \]
\end{lemma}
\end{subsubsection}

\begin{subsubsection}{{\bf A priori estimates for {\it Step 5}}} We next give the existence, uniqueness and regularity for the function
$\phi$ defined in {\it Step 5}. \newline

Consider Poisson's equation on the space-time sheath region
$\Lambda_s(T;\bv)$: Let $t \in [0, T]$ be given and $\phi$ satisfy
\begin{equation}
\begin{cases}
 \Delta \phi = n~~\mbox{ in } \Omega_s(t;\bv), \cr
 \nabla \phi \cdot \mbox{ \boldmath $\nu$}_0 = g  \quad \mbox{ on } \partial \Omega_0 \quad \mbox{ and } \quad
 \phi = -\ln n_s \quad \mbox{ on }{\mathcal S}(t).
\end{cases} \label{Ha1}
\end{equation}
%
%
\begin{lemma}
Let $n$ be an ion density in the sheath region $\Lambda_s(T;\bv)$
and satisfy the {\it a priori} estimate in Lemma A.8.  Then
Poisson's equation (\ref{Ha1}) has a unique solution $\phi$
satisfying the following estimate:
\[
\max_{ 1\leq |\alpha| \leq 2} |||\partial^{\alpha}
\phi|||_{0,\gamma, \bar \Lambda_s(T;\bv)}
 +  \sup_{0\leq t \leq T} \max_{|\alpha|=3} ||\partial^{\alpha} \phi(\cdot,t) )||_{0,\gamma, \bar \Omega_s(t;\bv)}  \leq R_3.
\]
Here $R_3$ is a positive constant only depending on $K_0,
\delta_{*i}, i=1,2$ and $\delta^*$ respectively.
\end{lemma}
\begin{proof}
(i) Differentiation of (\ref{Ha1}) with respect to $t$ shows that
$\partial_t \phi$ satisfies the mixed Dirichlet-Neumann problem
for Poisson's equation.
\[
\begin{cases}
 \Delta \partial_t \phi = -\mbox{div}(n\bv)~~\mbox{ in } \Omega_s(t;\bv), \cr
 \nabla \partial_t \phi \cdot \mbox{ \boldmath $\nu$}_0 = \partial_t g  \quad \mbox{ on } \partial \Omega_0 \quad \mbox{ and } \quad
 \partial_t \phi = - \partial_t \ln n_s \quad \mbox{ on }{\mathcal S}(t),
\end{cases} \label{Ha1-1}
\]
where we used $\nabla \phi \cdot \mbox{ \boldmath $\nu$} = 0~$ and
$~\nabla n \cdot \mbox{ \boldmath $\nu$} =0$ on the interface
${\mathcal S}(t),~0\leq t \leq T$.

By the direct application of H\"older estimates of the first
derivatives given in (\cite{G-T}, Section 8), we have
\[ || \partial_t \phi||_{2,\gamma, \bar \Omega_s(t;\bv)} \leq \bar C_1 \Big( ||\partial_t g||_{1,\gamma, \bar \Omega_s(t;\bv)} + |||n\bv|||_{1,\gamma, \bar \Omega_s(t;\bv)} +
|| \partial_t \ln n_s||_{2,\gamma, \bbr} \Big). \] Here $\bar C_1$
depends on  $\Omega_0$ and ${\mathcal S}(t)$, but we can choose
uniform $\bar C_1$ independent of $t$ and depending only on
$\delta_*$ and $\delta^*$ for sufficiently small $T$, $0\leq t
\leq T$.

On the other hand, since $\bv \in {\mathcal B}(T)$,  we have
\begin{eqnarray*}
&& \bullet~~|||n\bv|||_{1,\gamma, \bar \Omega_s(t;\bv)} \leq \bar
C_2||n||_{1,\gamma, \bar \Omega_s(t;\bv)} ||\bv||_{1,\gamma, \bar
\Omega_s(t;\bv)} \leq \bar C_3 K_0 (R_2 + R_3) \delta^*, \cr &&
\bullet~~||\partial_t g||_{1,\gamma, \bar \Omega_s(t;\bv)} \leq
\delta^* \quad \mbox{ by the assumption (A2) in Section 7.2},
\end{eqnarray*}
where $\bar C_2$ and $\bar C_3$ are some positive constants.

It follows from the interface equation (\ref{EQ7-2-8}) that
\begin{eqnarray*}
&& \displaystyle \frac{\partial_t n_s}{n_s} = -\Big(
\frac{4\sin\beta \sin \theta \theta}{r} \Big) \partial_{\beta}
\theta - \Big( \frac{2\sin \beta \tilde V \cos \theta}{rn_s} \Big)
\partial_{\beta}n_s, \cr && \tilde V = -1 - \frac{\nabla \zeta
\cdot (\cos\theta, \sin \theta)}{n_s}.
\end{eqnarray*}
We use the above relation and the estimates from Lemma A.5 (1) to
obtain
\[ || \partial_t \ln n_s||_{2,\gamma, \bbr} \leq C(\delta_{*1}, \delta^*). \]
Here $C(\delta_{*1}, \delta^*)$ is a positive constant depending
only on $\delta_{*1}, \delta^*$. Hence we have
\begin{equation}
\displaystyle \sup_{0 \leq t \leq T} || \partial_t
\phi(\cdot,t)||_{2,\gamma, \bar \Omega_s(t;\bv)} \leq R_{3,0}(K_0,
\delta_{*1}, \delta_{*2}, \delta^*) \quad \mbox{ for } t \in [0,
T]. \label{LA8-0}
\end{equation}
(ii) It follows from the Schauder estimates (Section 6.2 in
\cite{G-T}) that
\begin{eqnarray}
||\phi(\cdot,t)||_{2,\gamma, \bar \Omega_s(t;\bv)} &\leq& \bar C_4
\Big( ||n(\cdot,t)||_{0,\gamma, \bar \Omega_s(t;\bv)} +
||g(\cdot,t)||_{1,\gamma,\partial \Omega_0} + ||\ln n_s(\cdot,t)||_{2,\gamma,\bbr} \Big) \nonumber \\
&\leq& R_{3,1}(K_0, \delta_{*1}, \delta_{*2}, \delta^*), \quad 0
\leq t \leq T. \label{LA8-1}
\end{eqnarray}
Here $\bar C_4$ depends only on the $\Omega_0$ and ${\mathcal
S}(t)$, but again we can choose $\bar C_4$ depending only on
$\delta_*$ and $\delta^*$ for sufficiently small $T, 0 \leq t \leq
T$. \newline

Let $(\bx,t)$ and $(\by,s)$ be any points in $\Lambda_s(T;\bv)$.
Without loss of generality, we assume that $0\leq s < t$. By
assumption (A4) of Section 7.2, we have a contracting interface so
that
\[ \Omega_s(t;\bv) \subset \Omega_s(s;\bv), \quad  0 \leq s < t \leq T \ll 1. \]
Hence $\bx \in \Omega_s(t;\bv)$. Then inequality (\ref{LA8-0})
implies, for $\bx \in \Omega_s(t;\bv)$
\begin{equation}
 \max_{1\leq |\alpha| \leq 2} ||\partial^{\alpha} \phi(\bx,\cdot)||_{0,\gamma, [0,T]}
 \leq R_{3,0}(K_0, \delta_{*1}, \delta_{*2}, \delta^*) T^{1-\gamma}. \label{LA8-2}
\end{equation}
We combine (\ref{LA8-1}) and (\ref{LA8-2}) and choose $T$
sufficiently small to get
\begin{equation}
\max_{1\leq |\alpha| \leq 2} |||\partial^{\alpha}
\phi|||_{0,\gamma, \bar \Lambda_s(T;\bv)} \leq R_{3,2}(K_0,
\delta_{*1}, \delta_{*2}, \delta^*). \label{LA8-3}
\end{equation}
(In fact the above argument holds for the expanding interfaces as
well). \newline

(iii) On the other hand, $\partial_{x_i} \phi, i=1,2$ satisfies
\begin{equation}
\begin{cases}
 \Delta \partial_{x_i} \phi = \partial_{x_i} n~~\mbox{ in } \Omega_s(t;\bv), \cr
 \nabla \partial_{x_i} \phi \cdot \mbox{ \boldmath $\nu$}_0 = \partial_{x_i} g  \quad \mbox{ on } \partial \Omega_0 \quad \mbox{ and } \quad
 \partial_{x_i} \phi = -\partial_{x_i} \ln n_s \quad \mbox{ on }{\mathcal S}(t).
\end{cases} \label{Ha2}
\end{equation}
Again, it follows from the Poisson equation and the Schauder
estimates (Section 6.2 in \cite{G-T}) that
\begin{eqnarray}
&& ||\partial_{x_i} \phi(\cdot,t)||_{2,\gamma, \bar \Omega_s(t;\bv)} \nonumber \\
&& \leq \bar C_5 \Big( ||\partial_{x_i}
n(\cdot,t)||_{0,\gamma,\bar \Omega_s(t;\bv)} + ||\partial_{x_i}
g(\cdot,t)||_{1,\gamma,  \partial \Omega_0}+ ||\partial_{x_i} \ln
n_s(\cdot,t)||_{2,\gamma, \bbr} \Big). \label{LA8-3}
\end{eqnarray}
Here $\bar C_5$ depends on ${\mathcal S}(t)$, but we can choose
$\bar C_5$ depending only on $\delta_{*i}, i=1,2$ and $\delta^*$
for sufficiently small $T, 0 \leq t \leq T$. \newline

The first two terms in the right hand side of (\ref{LA8-3}) can be
bounded by a quantity depending on $\delta^*$ using Lemma A.7 and
assumptions (A1)-(A2) of the boundary data in Section 7.2, i.e.,
\begin{equation}
||\partial_{x_i} n(\cdot,t)||_{0,\gamma,\bar \Omega_s(t;\bv)} +
||\partial_{x_i} g(\cdot,t)||_{1,\gamma,  \partial \Omega_0} \leq
\bar C_6. \quad \label{LA8-4}
\end{equation}
Here $\bar C_6$ is a positive constant depending only on
$\delta_{*1}$ and $\delta^*$.

Now we estimate the third term $||\partial_{x_i} \ln
n_s||_{2,\gamma, \bbr}$ as follows. It follows from
(\ref{EQ7-2-4}) that we have
\[ \partial_{x_1} = -\frac{2\sin \beta}{r} \partial_{\beta} \]
and similarly we can express $\partial_{x_2}$ in terms of
$\partial_{\beta}$. Therefore we have
\begin{equation}
||\partial_{x_i} \ln n_s(\cdot,t)|||_{2,\gamma, \bbr} = \Big|
\Big|\frac{\partial_{x_i} n_s(\cdot,t)}{n_s(\cdot,t)} \Big|
\Big|_{2,\gamma, \bbr} \leq \bar C_7 \quad i=1,2. \label{LA8-5}
\end{equation}
Here $\bar C_7$ is a positive constant depending only on
$\delta_{*1}$ and $\delta^*$.

Combining estimates (\ref{LA8-4}) and (\ref{LA8-5}), we obtain
\[  \max_{|\alpha|=1} ||\partial^{\alpha} \phi(\cdot,t)||_{2,\gamma, \bar \Omega_s(t;\bv)} \leq R_{3,3}(K_0, \delta_{*1}, \delta_{*2}, \delta^*) \quad \mbox{ for } t \in [0,T]. \]
The above inequality implies
\[ \displaystyle  \sup_{0 \leq t \leq T} \max_{|\alpha|=1} ||\partial^{\alpha} \phi(\cdot,t)||_{0,\gamma, \bar \Omega_s(t;\bv)}
\leq R_{3,3}(K_0, \delta_{*1}, \delta_{*2}, \delta^*) \quad \mbox{
for } t \in [0,T]. \] In particular we have
\begin{equation}
\displaystyle  \sup_{0 \leq t \leq T} \max_{|\alpha|=3}
||\partial^{\alpha} \phi(\cdot,t)||_{0,\gamma, \bar
\Omega_s(t;\bv)} \leq R_{3,3}(K_0, \delta_{*1}, \delta_{*2},
\delta^*) \quad \mbox{ for } t \in [0,T]. \label{LA8-6}
\end{equation}
We set $R_3(K_0, \delta_{*1}, \delta_{*2} \delta^*) :=
R_{3,2}(K_0, \delta_{*1}, \delta_{*2}, \delta^*) +
R_{3,3}(K_0,\delta_{*1}, \delta_{*2}, \delta^*)$ and use
(\ref{LA8-3}) and (\ref{LA8-6}) to get the desired result.
\end{proof}

\end{subsubsection}

\begin{subsubsection}{{\bf A priori estimates for {\it Step 6}}} In this part, we give the existence, uniqueness and regularity for the ion velocity $\hat \bu$ defined
in {\it Step 6} of Section 7.2. \newline

Consider the Burgers' equation with a known source $\nabla \phi$:
\begin{equation}
\partial_t \hat \bu + ( \hat \bu \cdot \nabla) \hat \bu = \nabla \phi, \qquad (\bx,t) \in \bbr^2 \times \bbr_+. \label{PE}
\end{equation}

%
%
\begin{lemma}
Suppose the source $\nabla \phi$ satisfies the estimates obtained
in Lemma A.9. Also assume initial data $\bu_0$
 satisfy the assumption (A3) of Section 7.2 so that
\begin{enumerate}
\item $\nabla \phi \in C^{1,\gamma}(\bar \Lambda_s(T;\bv)) \quad
\mbox{ and } \quad \nabla \phi(\cdot,t) \in
C^{2,\gamma}(\Omega_s(t;\bv));$ \item for each $\mbox{\boldmath
$\alpha$} \in \bbr^2$,  the real parts of the eigenvalues of
$\nabla \bu_0(\mbox{\boldmath $\alpha$})$ are non-negative;
\end{enumerate}
the there is a positive constant $T_2$ such that (\ref{PE}) has a
unique solution $\hat \bu \in C^{1, \gamma}(\bar
\Lambda_s(T;\bv))$ satisfying
\begin{eqnarray}
&& \det \Gamma(\mbox{\boldmath $\alpha$},t) > 0 \quad \mbox{ and } \nonumber \\
&& \hat \bu(\mbox{\boldmath $\chi$}(t,0,\mbox{\boldmath
$\alpha$})) = \bu_0(\mbox{\boldmath $\alpha$}) + \int_0^t \nabla
\phi(\mbox{\boldmath $\chi$}(s,0,\mbox{\boldmath $\alpha$}),s)ds,
\quad t \in [0, T_2],\label{PE0} \\
&& \hat \bu(\bx,t) \cdot \bx \leq -\frac{\eta_0}{2} |\bx|^2, \quad
(\bx,t) \in (B(0,r_b + 6K_0 \delta^* T_2) - \Omega_0) \times
[0,T_2], \nonumber
\end{eqnarray}
where $T_2$ is a positive constant and
\[ \frac{d \mbox{\boldmath $\chi$}(t,0,\mbox{\boldmath $\alpha$})}{dt}  = \hat \bu(\mbox{\boldmath $\chi$}(t,0,\mbox{\boldmath $\alpha$}), t)
 \quad \mbox{ and } \quad \Gamma(\mbox{\boldmath $\alpha$},t) = \nabla  \hat \bu(\mbox{\boldmath $\alpha$},t). \]
\end{lemma}
\begin{proof}
(i) Along the particle path $\mbox{\boldmath
$\chi$}(t,0,\mbox{\boldmath $\alpha$})$, system (\ref{PE}) becomes
\begin{equation}
\displaystyle \frac{D \hat \bu}{Dt} =\nabla \phi, \quad \mbox{
where } \frac{D}{Dt} = \partial_t  + \hat \bu \cdot \nabla
\end{equation}
Any smooth solution of (\ref{PE0}) will satisfy
\begin{equation}
 \frac{d^2 \mbox{\boldmath $\chi$}(t,0,\mbox{\boldmath $\alpha$})}{dt^2}
 = \nabla \phi(\mbox{\boldmath $\chi$}(t,0,\mbox{\boldmath $\alpha$}),t); \qquad \mbox{\boldmath $\chi$}(\mbox{\boldmath $\alpha$},0)
  = \mbox{\boldmath $\alpha$}, \quad \frac{d\mbox{\boldmath $\chi$}(0,0,\mbox{\boldmath $\alpha$})}{dt} = \bu_0(\mbox{\boldmath $\alpha$}). \label{PE1}
\end{equation}
Since $\nabla \phi(\cdot,t)$ is Lipschitz continuous and uniformly
bounded, there exists a unique characteristic curve
$\mbox{\boldmath$\chi$}(t,0,\mbox{\boldmath $\alpha$})$ satisfying
(\ref{PE1}) locally in time $t$. Now we integrate (\ref{PE1}) to
get
\begin{equation}
\frac{d\mbox{\boldmath $\chi$}(t,0,\mbox{\boldmath $\alpha$})}{dt}
  = \bu_0(\mbox{\boldmath $\alpha$}) + \int_0^t \nabla \phi(\mbox{\boldmath $\chi$}(s,0,\mbox{\boldmath $\alpha$}),s)ds. \label{PE2}
\end{equation}
and  integration of the above equation yields
\begin{eqnarray}
\mbox{\boldmath $\chi$}(t,0,\mbox{\boldmath $\alpha$}) &=&
\mbox{\boldmath $\chi$}(0,0,\mbox{\boldmath $\alpha$})
+ t \bu_0(\mbox{\boldmath $\alpha$})  + \int_0^t \int_0^{t_1} \nabla \phi(\mbox{\boldmath $\chi$}(s,0,\mbox{\boldmath $\alpha$}),s)dsdt_1 \nonumber \\
 &=& \mbox{\boldmath $\alpha$} + t \bu_0(\mbox{\boldmath $\alpha$})
 + \int_0^t \int_0^{t_1} \nabla \phi(\mbox{\boldmath $\chi$}(s,0,\mbox{\boldmath $\alpha$}),s)dsdt_1. \label{PE3}
\end{eqnarray}
Next we differentiate (\ref{PE3}) with respect to $\mbox{\boldmath
$\alpha$}$ to get
\begin{equation}
 \Gamma(\mbox{\boldmath $\alpha$}, t) = I +  t\nabla \bu_0(\mbox{\boldmath $\alpha$})
 + \int_0^t \int_0^{t_1} (\nabla \otimes \nabla) \phi (\mbox{\boldmath $\chi$}(s,0,\mbox{\boldmath $\alpha$}),s)
 \Gamma(\mbox{\boldmath $\alpha$},s)dsdt_1. \label{NPE}
\end{equation}
Set
\[ y(t) =| \Gamma(\mbox{\boldmath $\alpha$}, t) -I -  t\nabla \bu_0(\mbox{\boldmath $\alpha$})|, \]
where $|\cdot|$ denotes any norm on $2\times 2$ matrices so that
we have
\[ y(t) \leq \int_0^t \int_0^{t_1} |(\nabla \otimes \nabla) \phi (\mbox{\boldmath $\chi$}(s,0,\mbox{\boldmath $\alpha$}),s)|
 |\Gamma(\mbox{\boldmath $\alpha$},s)|dsdt_1. \]
 Since
 \begin{eqnarray*}
  |\Gamma(\mbox{\boldmath $\alpha$}, t)| &=& |\Gamma(\mbox{\boldmath $\alpha$}, t) -I -  t\nabla \bu_0(\mbox{\boldmath $\alpha$})|
 + |I +  t\nabla \bu_0(\mbox{\boldmath $\alpha$})| \cr
  &\leq&  y(t) + 1 + d_1 t,
  \end{eqnarray*}
  where $\displaystyle \sup_{\mbox{\boldmath $\alpha$} \in \Omega(0)} |\nabla \bu_0(\mbox{\boldmath $\alpha$})| \leq d_1$, we have from (\ref{NPE}) that
\[
y(t) \leq \int_0^t \int_0^{t_1} d_2 (y(s) + 1 + d_1 s)dsdt_1,
\]
where $|\nabla \otimes \nabla \phi(\mbox{\boldmath
$\chi$}(s,0,\mbox{\boldmath $\alpha$})| \leq d_2$ for
$\mbox{\boldmath $\alpha$} \in \Omega(0), 0 \leq s\leq T$. Hence
we have
\[ y(t) \leq d_2 \Big( \frac{t^2}{2} + d_1 \frac{t^3}{3} \Big) + d_2 \int_0^t \int_0^{t_1} y(s)dsdt_1 \]
and by Appendix B
\[ y(t) \leq d_3 t^2 \qquad \mbox{ on } 0 \leq t \leq T, \]
for sufficiently small $T$, i.e.,
\[ |\Gamma(\mbox{\boldmath $\alpha$}, t) -I -  t\nabla \bu_0(\mbox{\boldmath $\alpha$})| \leq d_3 t^2, \quad 0 \leq t \leq T. \]
Define
\[ tD(t, \mbox{\boldmath $\alpha$}) := \Gamma(\mbox{\boldmath $\alpha$}, t) -I -  t\nabla \bu_0(\mbox{\boldmath $\alpha$}). \]
Then we have
\[ |D(t, \mbox{\boldmath $\alpha$})| \leq d_3 t \quad \mbox{ for some constant $d_3 >0$} \]
and
\[ \mbox{det}(\Gamma(\mbox{\boldmath $\alpha$}, t)) = \mbox{det}( I + t(\nabla \bu_0(\mbox{\boldmath $\alpha$}) + D(t, \mbox{\boldmath $\alpha$}))). \]
Let $\lambda_i(\mbox{\boldmath $\alpha$}, t)$ and
$\lambda_i(\mbox{\boldmath $\alpha$}), i=1,2$ be the eigenvalues
of a matrix $\nabla \bu_0(\mbox{\boldmath $\alpha$}) + D(t,
\mbox{\boldmath $\alpha$})$ \newline and $\nabla
\bu_0(\mbox{\boldmath $\alpha$})$ respectively. Then we can see
that
\[ \lambda_i(\mbox{\boldmath $\alpha$}, t) = \lambda_i(\mbox{\boldmath $\alpha$}) + {\mathcal O}(t). \]
By the Cayley-Hamilton theorem, we have
\begin{eqnarray*}
\mbox{det}(\Gamma(\mbox{\boldmath $\alpha$}, t)) &=&
(t\lambda_1(\mbox{\boldmath $\alpha$},t) +
1)(t\lambda_2(\mbox{\boldmath $\alpha$},t) + 1) \cr &=&
(t\lambda_1(\mbox{\boldmath $\alpha$}) +
1)(t\lambda_2(\mbox{\boldmath $\alpha$}) + 1) + {\mathcal O}(t^2)
\cr &=& \mbox{det}(I + t \nabla \bu_0(\mbox{\boldmath $\alpha$}))
+ {\mathcal O}(t^2).
\end{eqnarray*}
As long as $0 \leq t \leq T \ll1$, the sign of
$\mbox{det}(\Gamma(\mbox{\boldmath $\alpha$}, t)) $ will be
determined by $\det\Big(I + t \nabla \bu_0(\mbox{\boldmath
$\alpha$})\Big)$. \newline

Next we calculate $\det ( I + t \nabla \bu_0(\mbox{\boldmath
$\alpha$}))$. Let us set the characteristic polynomial of $\nabla
\bu_0(\mbox{\boldmath $\alpha$})$ by $P(\mbox{\boldmath $\alpha$},
\lambda)$. Then we have
\[
 P(\mbox{\boldmath $\alpha$}, \lambda) \equiv \det (\nabla \bu_0(\mbox{\boldmath $\alpha$}) - \lambda I)
 = (\lambda_1(\mbox{\boldmath $\alpha$}) - \lambda) (\lambda_2(\mbox{\boldmath $\alpha$}) - \lambda),\]
where $\lambda_i(\mbox{\boldmath $\alpha$})$ are the eigenvalues
of $\nabla \bu_0(\mbox{\boldmath $\alpha$})$. Hence
\begin{eqnarray*}
\det\Big(I + t \nabla \bu_0(\mbox{\boldmath $\alpha$})\Big)  &=&
t^2 \det \Big( \nabla \bu_0(\mbox{\boldmath $\alpha$}) + t^{-1} I
\Big) = t^2 P(\mbox{\boldmath $\alpha$}, -t^{-1}) \cr &=& t^2
(\lambda_1(\mbox{\boldmath $\alpha$}) + t^{-1})
(\lambda_2(\mbox{\boldmath $\alpha$}) + t^{-1}) \cr
&=&(t\lambda_1(\mbox{\boldmath $\alpha$}) +
1)(t\lambda_2(\mbox{\boldmath $\alpha$}) + 1).
\end{eqnarray*}
Since by assumption (2) above, real parts of the eigenvalues of
the Jacobian matrix $\nabla \bu_0(\mbox{\boldmath $\alpha$})$  are
nonnegative, we have
\[ \displaystyle \det \Gamma(\mbox{\boldmath $\alpha$}, t) \geq  \Pi_{q =1}^{2} [1 +
t \mbox{ Re } \lambda_q (\mbox{\boldmath $\alpha$})]  + {\mathcal
O}(t^2) > 0, \quad 0 \leq t \leq  T \ll 1. \] Hence the Lagrangian
map is a $C^1$-diffeomorphism locally in time. \newline

\noindent (ii)  It follows from (\ref{PE0}) that
\[ \hat \bu(\mbox{\boldmath $\chi$}(t,0,\mbox{\boldmath $\alpha$}),t) = \bu_0(\mbox{\boldmath $\alpha$}) + \int_0^t \nabla
\phi(\mbox{\boldmath $\chi$}(s,0,\mbox{\boldmath $\alpha$}),s)ds.
\] We take an inner product with $\mbox{\boldmath
$\chi$}(t,0,\mbox{\boldmath $\alpha$}),t)$ to get
\begin{eqnarray}
&& \langle \hat \bu(\mbox{\boldmath $\chi$}(t,0,\mbox{\boldmath $\alpha$}), t), \mbox{\boldmath $\chi$}(t,0,\mbox{\boldmath $\alpha$})  \rangle \nonumber \\
&& = \langle \bu_0(\mbox{\boldmath $\alpha$}), \mbox{\boldmath
$\chi$}(t,0,\mbox{\boldmath $\alpha$}) \rangle + \int_0^t  \langle
\nabla
\phi(\mbox{\boldmath $\chi$}(s,0,\mbox{\boldmath $\alpha$}),  \mbox{\boldmath $\chi$}(t,0,\mbox{\boldmath $\alpha$}) \rangle ds \nonumber \\
&& = \langle \bu_0(\mbox{\boldmath $\alpha$}), \mbox{\boldmath
$\chi$}(t,0,\mbox{\boldmath $\alpha$}) - \mbox{\boldmath $\alpha$}
\rangle + \langle \bu_0(\mbox{\boldmath $\alpha$}),
\mbox{\boldmath $\alpha$}  \rangle  + \int_0^t  \langle \nabla
\phi(\mbox{\boldmath $\chi$}(s,0,\mbox{\boldmath $\alpha$}),
\mbox{\boldmath $\chi$}(t,0,\mbox{\boldmath $\alpha$}) \rangle ds.
\label{NI}
\end{eqnarray}
Since
\begin{eqnarray*}
&& |\mbox{\boldmath $\chi$}(t,0,\mbox{\boldmath $\alpha$}) -
\mbox{\boldmath $\alpha$}| = \Big| \int_0^t \bv(\mbox{\boldmath
$\chi$}(s,0,\mbox{\boldmath $\alpha$}),s)ds \Big| \leq 6K_0
\delta^* T, \cr && \mbox{ and } \quad |||\nabla \phi|||_{0, \bar
\Lambda_s(T;\bv)} \leq R_3,
\end{eqnarray*}
Hence in (\ref{NI}), we have
\[
\langle \hat \bu(\mbox{\boldmath $\chi$}(\mbox{\boldmath
$\alpha$}, t), t), \mbox{\boldmath $\chi$}(\mbox{\boldmath
$\alpha$},t)  \rangle  \leq 6K_0 (\delta^*)^2 T -\eta_0
||\mbox{\boldmath $\alpha$}||^2
 + R_3 T.
\]
Now we choose $T$ sufficiently small so that
\[ \Big( 6K_0 (\delta^*)^2 + R_3 \Big) T \leq  \frac{\eta_0 r_a^2}{2} \leq   \frac{\eta_0}{2}||\mbox{\boldmath $\alpha$}||^2, \qquad  \mbox{\boldmath $\alpha$} \in
\Omega_s(0). \] Then we have
\[ \langle \hat \bu(\mbox{\boldmath $\chi$}(\mbox{\boldmath $\alpha$},
t), t), \mbox{\boldmath $\chi$}(\mbox{\boldmath $\alpha$},t)
\rangle  \leq -\frac{\eta_0}{2}||\mbox{\boldmath $\alpha$}||^2. \]
On the other hand, since
\[
\frac{d}{ds}|\mbox{\boldmath $\chi$}(0,0, \mbox{\boldmath
$\alpha$})|^2 \leq -2 \eta_0 |\mbox{\boldmath $\chi$}(0,0,
\mbox{\boldmath $\alpha$})|^2 = -2 \eta_0 |\mbox{\boldmath
$\alpha$}|^2,
\]
we have
\[ |\mbox{\boldmath $\chi$}(t ,0, \mbox{\boldmath $\alpha$})| \leq |\mbox{\boldmath $\alpha$}| \quad \mbox{ for } t \leq T \ll 1. \]
Therefore we obtain
\[ \langle \hat \bu(\mbox{\boldmath $\chi$}(t,0,\mbox{\boldmath $\alpha$}), t), \mbox{\boldmath $\chi$}(t,0,\mbox{\boldmath $\alpha$})  \rangle
 \leq -\frac{\eta_0}{2} |\mbox{\boldmath $\chi$}(t ,0, \mbox{\boldmath $\alpha$})|^2. \]
\end{proof}

\end{subsubsection}

\begin{subsubsection}{{\bf A priori estimates for {\it Step 7} of Section 7.2} }
Finally we prove the existence of a linear extension map and some
estimates of the extension.
%
%
\begin{lemma}
Let ${\mathcal S}(t), t \in [0, T] $ be the $C^{2,\gamma}$-regular
simple closed convex curve in $\bbr^2$ provided by Lemma A.6 such
that ${\mathcal S}(t)$ lies inside the annulus $\Omega_*$ and
$\Omega_s(t;\bv)$ is the corresponding sheath region $0 \leq t
\leq T$, T sufficiently small. Then there exists a bounded linear
operator ${\mathcal E}(\cdot; {\mathcal S}(t)):
C^{2,\gamma}(\Omega_s(t;\bv)) \to C^{2,\gamma}(\Omega_1)$
satisfying
 \begin{eqnarray*}
&& \mbox{(a)}~{\mathcal E}( \hat \bu| {\mathcal S}(t)) = \hat \bu,
~\mbox{ in } \Omega_s(t;\bv), \cr && \mbox{(b)}~{\mathcal E}( \hat
\bu| {\mathcal S}(t)) \mbox{ has support in } B(0, 3\delta^*), \cr
&& \mbox{(c)}~|||{\mathcal E}(\hat \bu| {\mathcal
S}(t))|||_{2,\gamma,\bar \Omega_1} \leq K_0 |||\hat
\bu|||_{2,\gamma, \bar \Omega_s(t;\bv)},
\end{eqnarray*}
where $K_0$ is independent of $t \in [0, T]$ and $\Omega_*$ is the
annulus region (\ref{Set1}).
\end{lemma}
\begin{proof}
Since the proof is rather long, we delay its proof until Appendix
C.
\end{proof}

\end{subsubsection}
\end{subsection}

\begin{subsection}{{\bf Continuity of ${\mathcal F}$}} In this part, we establish the continuity of ${\mathcal F}$
which in turn imply the existence of a fixed point of ${\mathcal
F}$.
%
%
\begin{lemma}
Let $f$ be a continuous function such that
\begin{eqnarray*}
&& \displaystyle f(t) \leq C_0 + C_1 (f(t))^2, \quad t \geq 0, \cr
&& \displaystyle f(0) \leq C_0 \quad \mbox{ and } \quad C_0C_1
\leq \frac{1}{8},
\end{eqnarray*}
where $C_0$ and $C_1$ are positive constants independent of $t$.
Then we have
\[\displaystyle  f(t) \leq 2 C_0. \]
\end{lemma}
\begin{proof}
Define
\[ F(k) = C_1 k^2 - k + C_0. \]
Then by direct calculation, we have
\[ \displaystyle  \min F(k) = \frac{-1+ 2 C_0C_1}{2C_1} <0 \quad \mbox{ at } k = \frac{1}{2C_1}. \]
Now we denote $r_1$ and $r_2$ by the roots of $F(k) = 0$ such that
$r_1 < r_2$. Then  by direct calculation, the smallest root $r_1$
satisfies
\[ C_0 \leq r_1 = \frac{2C_0}{1 + \sqrt{1-4C_0C_1}} \leq 4(\sqrt{2} -1) C_0 \leq 2C_0. \]

On the other hand, since $F(f(t)) \geq 0$, we have two cases:
\[ \mbox{ either } f \leq r_1 \quad \mbox{ or } \quad f \geq  r_2, \]
however since $f(0) \leq C_0 \leq r_1$ and $f(t)$ is continuous,
we have
\[ \displaystyle  f \leq r_1  \leq  2C_0. \]
\end{proof}
\begin{proposition}
There exists a positive constant $T$ such that the map ${\mathcal
F}$ with ${\mathcal F}(\bv) := {\mathcal E}(\hat \bu; {\mathcal
S}(t))$ is a well-defined map from ${\mathcal B}(T)$ to ${\mathcal
B}(T)$.
\end{proposition}
\begin{proof} For the time being, we assume $T$ sufficiently small so that
\begin{equation}
T \leq \min\{ T_1, T_2\}. \label{P7.1-1}
\end{equation}
So all estimates in the previous lemmas hold. \newline (i) By the
construction of $\hat \bu$ in the sheath region
$\Lambda_s(T;\bv)$, we have from solving (\ref{PE}) along the
characteristic
\begin{equation}
\displaystyle  \hat \bu(\bx,t) =
\begin{cases}
\bu_0(\mbox{\boldmath $\alpha$}(\bx,t)) + \int_0^t \nabla
\phi(\mbox{\boldmath $\hat \chi$}(s,t,\bx),s)ds ~~& t_0 = 0, \cr
-\mbox{\boldmath $\nu$}(\mbox{\boldmath $\alpha$}(\bx,t)) +
\int_{t_0}^t \nabla \phi(\mbox{\boldmath $\hat
\chi$}(s,t,\bx),s)ds ~~& t_0 > 0. \label{P7.1-2}
\end{cases}
\end{equation}
In (\ref{P7.1-2}), we have
\begin{eqnarray*}
\displaystyle ||\hat u_i||_{0, \bar \Lambda_s(T;\bv)} &\leq&
\begin{cases}
\displaystyle ||u_{i0}||_{0, \bar \Omega_s(0)} + t |||\nabla
\phi|||_{0, \bar \Lambda_s(T;\bv)}~~& t_0 = 0, \cr 1 + (t-t_0)
|||\nabla \phi|||_{0,\bar \Lambda_s(T;\bv)}~~& t_0 > 0.
\end{cases}
\end{eqnarray*}
This of course implies
\begin{eqnarray}
||\hat u_i||_{0,\bar \Lambda_s(T;\bv)} &\leq&  || u_{i0}||_{0,\bar \Omega_s(0)} + T|||\nabla \phi|||_{0,\bar \Lambda_s(T;\bv)} \nonumber \\
   &\leq& ||u_{0i}||_{0,\bar \Omega_s(0)} + TR_3 \qquad \mbox{ by Lemma A.8}. \label{P7.1-3}
\end{eqnarray}
On the other hand, we use Lemma A.2 to obtain
\begin{equation}
[[\hat u_i]]_{0,\gamma, \bar \Lambda_s(T;\bv)} \leq
[u_{0i}]_{0,\gamma, \bar  \Omega_s(0)} + C_1(T) R_3.
\label{P7.1-4}
\end{equation}
We combine (\ref{P7.1-3}) and (\ref{P7.1-4}) to get
\begin{equation}
||\hat u_i||_{0,\gamma, \bar \Lambda_s(T;\bv)} \leq
||u_{0i}||_{0,\gamma, \bar \Omega_s(0)} +  TR_3 + C_1(T)R_3.
\label{P7.1-4-1}
\end{equation}
We assume $T$ sufficiently small so that
\begin{equation}
TR_3 + C_1(T)R_3 \leq \frac{\delta^*}{3}. \label{P7.1-5}
\end{equation}
Here we used $C_1(T) ={\mathcal O}(T^{1-\gamma})$. Hence we have
from (\ref{P7.1-5}) that
\begin{equation}
\displaystyle \max_{i=1,2} || \hat u_i|||_{0,\gamma, \bar
\Lambda_s(T;\bv)} \leq \max_{i=1,2} ||u_{0i}||_{0,\gamma, \bar
\Omega_s(0)} + \frac{\delta^*}{3}. \label{P7.1-6}
\end{equation}
\noindent (ii) We differentiate the momentum equation
\[ \partial_t \hat \bu + \hat  \bu \cdot \nabla \hat \bu = \nabla \phi \]
with respect to $x_k$ to find
\begin{equation}
\frac{D (\partial_{x_k}\hat u_i)}{Dt} + \sum_{j=1}^{3}
(\partial_{x_k} \hat u_j) (\partial_{x_i} \hat u_j) =
\partial_{x_k} (\partial_{x_i} \phi), \label{P7.1-7}
\end{equation}
where $\frac{D}{Dt} = \partial_t + \hat \bu \cdot \nabla$. We
integrate (\ref{P7.1-7}) along the characteristic to get
\begin{eqnarray}
\displaystyle && (i) \mbox{ if } t_0 = 0, \nonumber \\
\displaystyle && \partial_{x_k} \hat u_i(\bx,t) = \partial_{x_k}
u_{0i}(\mbox{\boldmath $\alpha$}(\bx,t)) - \sum_{j=1}^{2} \int_0^t
\Big((\partial_{x_k} \hat u_j) (\partial_{x_i} \hat u_j)
\Big)(\mbox{\boldmath $\hat \chi$}(s,t,\bx),s)ds \cr
\displaystyle  &&  \hspace{3cm} + \int_0^t \partial_{x_k} (\partial_{x_i}\phi)(\mbox{\boldmath $\hat \chi$}(s,t,\bx),s)ds; \nonumber \\
\displaystyle && (ii) \mbox{  and if } t_0 >0, \cr \displaystyle
&& \partial_{x_k} \hat u_i(\bx,t) = -\partial_{x_k}
\nu_i(\mbox{\boldmath $\alpha$}(\bx,t)) -\sum_{j=1}^{2}
\int_{t_0}^t \Big((\partial_{x_k} \hat u_j)
(\partial_{x_i} \hat u_j) \Big)(\mbox{\boldmath $\hat \chi$}(s,t,\bx),s)ds \nonumber \\
\displaystyle && \hspace{3cm} + \int_{t_0}^t
\partial_{x_k}(\partial_{x_i} \phi)(\mbox{\boldmath $\hat
\chi$}(s,t,\bx),s)ds. \label{NP7.1}
\end{eqnarray}
The above equalities yield
\begin{eqnarray*}
\max_{i=1,2} \max_{|\alpha|=1} ||| \partial^{\alpha} \hat
u_i|||_{0, \bar \Lambda_s(T;\bv)} &\leq&  \max_{i=1,2}
\max_{|\alpha|=1} ||\partial^{\alpha} u_{0i}||_{0,\bar
\Omega_s(0)} + 2 T \Big(\max_{i=1,2} \max_{|\alpha|=1} |||
\partial^{\alpha} \hat u_i|||_{0, \bar \Lambda_s(T;\bv)} \Big)^2
\cr &+& T  \max_{|\alpha|=2}  ||| \partial^{\alpha} \phi
|||_{0,\bar \Lambda_s(T;\bv)}, \quad 0 \leq t \leq T.
\end{eqnarray*}
Since $T \ll 1$, it follows from Lemma A.12 that
\begin{eqnarray}
\max_{i=1,2}\max_{|\alpha|=1} ||| \partial^{\alpha} \hat
u_i|||_{0, \bar \Lambda_s(T;\bv)}  &\leq& 2 \Big(\max_{i=1,2}
\max_{|\alpha|=1} |||\partial^{\alpha} u_{0i}||_{0,\bar
\Omega_s(0)}
+ T \max_{|\alpha|=2}  ||| \partial^{\alpha} \phi |||_{0,\bar \Lambda_s(T;\bv)} \Big) \nonumber \\
 &\leq& 2 \max_{i=1,2}\max_{|\alpha|=1} ||\partial^{\alpha} u_{0i}||_{0,\bar \Omega_s(0)}  + 2T R_3. \label{P7.1-9}
\end{eqnarray}
On the other hand, it follows from the inequalities (\ref{NP7.1})
and (\ref{H1}) that
\begin{eqnarray*}
&& \displaystyle  \max_{i=1,2} \max_{|\alpha|=1}
[[\partial^{\alpha} \hat u_i]]_{0,\gamma,\bar \Lambda_s(T;\bv)}
\cr
 && \qquad \leq \max_{i=1,2} \max_{|\alpha|=1}  [\partial^{\alpha} u_{0i}]_{0,\gamma,\bar \Omega_s(0)}
+  4C_1(T) \Big( 2 \max_{i=1,2}
\max_{|\alpha|=1}||\partial^{\alpha} \hat u_{0i}||_{0,\bar
\Omega_s(0)} + 2TR_3 \Big) \cr && \qquad \times
\max_{i=1,2}\max_{|\alpha|=1}  [[\partial^{\alpha}\hat
u_i]]_{0,\gamma, \bar \Lambda_s(T;\bv)}
 + C_1(T)R_3.
\end{eqnarray*}
We assume  that
\[ 4C_1(T) \Big( 2 \max_{i=1,2}\max_{|\alpha|=1} ||\partial^{\alpha} u_{0i}||_{0,\bar \Lambda_s(T;\bv)} + 2 TR_3 \Big) \leq \frac{1}{2}. \]
Here we used $C_1(T) = {\mathcal O}(T^{1-\gamma})$. \newline

 Then we have
\begin{equation}
\displaystyle  \max_{i=1,2} \max_{|\alpha|=1} [[\partial^{\alpha}
\hat u_i]]_{0,\gamma,\bar \Lambda_s(T;\bv)} \leq 2 \max_{i=1,2}
\max_{|\alpha|=1}  [[\nabla u_{0i}]]_{0,\gamma,\bar \Omega_s(0)} +
2 C_1(T)R_3. \label{P7.1-10}
\end{equation}
We combine (\ref{P7.1-9}) and (\ref{P7.1-10}) to get
\[
\max_{i=1,2} \max_{|\alpha|=1}  |||\partial^{\alpha} \hat
u_i|||_{0,\gamma,\bar \Lambda_s(T;\bv)} \leq 2 \max_{i=1,2}
\max_{|\alpha|=1}  ||\partial^{\alpha} u_{0i}||_{0,\gamma,\bar
\Omega_s(0)} + 2 TR_3 + 2 C_1(T)R_3.
\]
We assume again that $T$ is sufficiently small so that
\begin{equation}
 2TR_3 + 2 C_1(T)R_3 \leq \frac{\delta^*}{3}. \label{P7.1-11}
\end{equation}
Then we have
\begin{equation}
\max_{i=1,2} |||\nabla \hat u_i|||_{0,\gamma,\bar
\Lambda_s(T;\bv)} \leq 2 \max_{i=1,2} ||\nabla
u_{0i}||_{0,\gamma,\bar \Omega_s(0)} + \frac{\delta^*}{3}.
\label{P7.1-12}
\end{equation}
(iii) We differentiate (\ref{P7.1-7}) with respect to $x_l$ to
obtain
\begin{eqnarray}
&& \frac{D(\partial_{x_k x_l}^2 \hat u_i)}{Dt} + \sum_{j=1}^{3}
\Big[ (\partial_{x_l} \hat u_j)(\partial_{x_jx_k}^2 \hat u_i) +
(\partial_{x_i} \hat u_j)
(\partial_{x_kx_l}^2 \hat u_j) + (\partial_{x_k} \hat u_j) (\partial_{x_ix_l}^2 \hat u_j) \Big] \nonumber \\
&& \qquad \qquad = \partial_{x_kx_l}^2 (\partial_{x_i} \phi).
\label{P7.1-13}
\end{eqnarray}
We integrate the equation (\ref{P7.1-13}) along the characteristic
curve to find
\begin{eqnarray}
\displaystyle && (i) \mbox{ if } t_0 = 0, \nonumber \\
\displaystyle && \partial_{x_k x_l}^2 \hat u_i(\bx,t) = \partial_{x_k x_l}^2 u_{0i}(\mbox{\boldmath $\alpha$}(\bx,t)) \nonumber \\
\displaystyle &&  \qquad -\sum_{j=1}^{2} \int_0^t
\Big((\partial_{x_l} \hat u_j)(\partial_{x_jx_k}^2 \hat u_i) +
(\partial_{x_i} \hat u_j)(\partial_{x_kx_l}^2 \hat u_j) +
(\partial_{x_k} \hat u_j) (\partial_{x_ix_l}^2 \hat u_j)
\Big)(\mbox{\boldmath $\hat \chi$}(s,t,\bx),s)ds \nonumber \\
\displaystyle && \qquad  + \int_0^t \partial_{x_k}\partial_{x_l} (\partial_{x_i} \phi)(\mbox{\boldmath $\hat \chi$}(s,t,\bx),s)ds; \nonumber \\
\displaystyle && (ii) \mbox{ if } t_0 >0, \nonumber \\
\displaystyle && \partial_{x_k} \partial_{x_l} \hat u_i(\bx,t) = -\partial_{x_k x_l}^2 \nu_i(\mbox{\boldmath $\alpha$}(\bx,t)) \nonumber \\
\displaystyle && \qquad -\sum_{j=1}^{2} \int_{t_0}^t
\Big((\partial_{x_l} \hat u_j)(\partial_{x_jx_k}^2 \hat u_i) +
(\partial_{x_i} \hat u_j)(\partial_{x_kx_l}^2 \hat u_j)
+ (\partial_{x_k} \hat u_j) (\partial_{x_ix_l}^2 \hat u_j) \Big)(\mbox{\boldmath $\hat \chi$}(s,t,\bx),s)ds \nonumber \\
\displaystyle && \qquad + \int_{t_0}^t \partial_{x_kx_l}^2 (\partial_{x_i} \phi)(\mbox{\boldmath $\hat \chi$}(s,t,\bx),s)ds. \nonumber \\
\label{P7.1-14}
\end{eqnarray}
Then it follows from (\ref{P7.1-14}) that
\begin{eqnarray*}
\displaystyle && \max_{i=1,2} \max_{|\alpha|=2}
|||\partial^{\alpha} \hat u_i|||_{0,\bar \Lambda_s(T;\bv)} \cr
 && \qquad \leq \max_{i=1,2} \max_{|\alpha|=2}||\partial^{\alpha} \hat u_{0i}||_{0, \bar \Omega_s(0)} + 6 T
 \Big( 2 \max_{i=1,2} \max_{|\alpha|=1}  ||\partial^{\alpha} u_{0i}||_{0,\gamma,\bar \Omega_s(0)} + \frac{\delta^*}{3} \Big) \cr
 &&  \qquad \times \max_{i=1,2} \max_{|\alpha|=2} |||\partial^{\alpha} \hat u_i|||_{0,\bar \Lambda_s(T;\bv)} + T R_3.
\end{eqnarray*}
We choose $T$ sufficiently small so that
\begin{equation}
6 T \Big( 2 \max_{i=1,2} \max_{|\alpha|=1}  ||\partial^{\alpha}
u_{0i}||_{0,\gamma,\bar \Omega_s(0)} + \frac{\delta^*}{3} \Big)
\leq \frac{1}{2}
 \quad \mbox{ and } \quad T R_3 \leq \frac{\delta^*}{12}. \label{P7.1-15}
\end{equation}
Then we have
\begin{equation}
\displaystyle \max_{i=1,2} \max_{|\alpha|=2} |||\partial^{\alpha}
\hat u_i|||_{0,\bar \Lambda_s(T;\bv)} \leq 2 \max_{i=1,2}
\max_{|\alpha|=2}||\partial^{\alpha}  u_{0i}||_{0, \bar
\Omega_s(0)} + \frac{\delta^*}{6}. \label{P7.1-16}
\end{equation}
We need to check the H\"older seminorm of $\partial^{\alpha} \hat
u_i$. Again we use (\ref{P7.1-14}) to find
\begin{eqnarray*}
&& \displaystyle \max_{i=1,2} \max_{|\alpha|=2}
[[\partial^{\alpha} \hat u_i]]_{0,\gamma, \bar \Lambda_s(T;\bv)}
\cr && \qquad \leq \max_{i=1,2} [[\partial^{\alpha}
u_{0i}]]_{0,\gamma, \bar \Lambda_s(T;\bv)} + 6  C_1(T) \Big( 2
\max_{i=1,2}  \max_{|\alpha|=1}  ||\partial^{\alpha}
u_{0i}||_{0,\gamma,\bar \Omega_s(0)} + \frac{\delta^*}{3} \Big)
\cr
 && \qquad \times \max_{i=1,2}  \max_{|\alpha|=2}  [[\partial^{\alpha} \hat u_i]]_{0,\gamma, \bar \Lambda_s(T;\bv)} + 6 C_1(T) \Big( 2 \max_{i=1,2}
 \max_{|\alpha|=1} ||\partial^{\alpha} u_{0i}||_{0,\gamma,\bar \Omega_s(0)} + \frac{\delta^*}{3} \Big) \cr
 && \qquad  \times \Big( 2\max_{i=1,2}\max_{|\alpha|=2} ||\partial^{\alpha}  u_{0i}||_{0, \bar \Omega_s(0)} + \frac{\delta^*}{6} \Big) +  C_1(T) R_3.
\end{eqnarray*}
Here we have used (\ref{H1}).

We assume that $T$ sufficiently is sufficiently small that
\begin{eqnarray}
&& 6 C_1(T) \Big( 2 \max_{i=1,2} \max_{|\alpha|=1}
||\partial^{\alpha} u_{0i}||_{0,\gamma,\bar \Omega_s(0)} +
\frac{\delta^*}{3} \Big) \leq \frac{1}{2}
 \quad \mbox{ and }
\label{P7.1-17} \\
&& 6 C_1(T) \Big( 2 \max_{i=1,2} \max_{|\alpha|=1}  ||\partial^{\alpha} u_{0i}||_{0,\gamma,\bar \Omega_s(0)} + \frac{\delta^*}{3} \Big) \nonumber \\
&& \times \Big( 2\max_{i=1,2} \max_{|\alpha|=2}
||\partial^{\alpha} \hat u_{0i}||_{0,\Omega_s(0)} +
\frac{\delta^*}{6} \Big)
 +  C_1(T) R_3 \leq \frac{\delta^*}{12}. \label{P7.1-18}
\end{eqnarray}
Hence we have
\begin{equation}
 \displaystyle \max_{i=1,2} \max_{|\alpha|=2} [[\partial^{\alpha} \hat u_i]]_{0,\gamma, \bar \Lambda_s(T;\bv)} \leq 2\max_{i=1,2}
 \max_{|\alpha|=2}  [\partial^{\alpha} u_{0i}]_{0,\gamma, \bar \Omega_s(0)} + \frac{\delta^*}{6}. \label{P7.1-19}
\end{equation}
We combine (\ref{P7.1-16}) and (\ref{P7.1-19}) to get
\begin{equation}
\displaystyle \max_{i=1,2} \max_{|\alpha|=2} |||\partial^{\alpha}
\hat u_i |||_{0,\gamma,\bar \Lambda_s(T;\bv)} \leq 2 \max_{i=1,2}
\max_{|\alpha|=2}  ||\partial^{\alpha}
 u_{0i} ||_{0,\gamma,\bar \Omega_s(0)} + \frac{\delta^*}{3}. \label{P7.1-20}
\end{equation}
We combine (\ref{P7.1-6}), (\ref{P7.1-12}) and (\ref{P7.1-20}) to
get
\begin{equation}
\displaystyle \max_{i=1,2}  \sum_{0 \leq k \leq 2} \max_{|\alpha|
= k}|||\partial^{\alpha}  \hat u_i |||_{0,\gamma,\bar
\Lambda_s(T;\bv)}
 \leq 2 \max_{i=1,2} \sum_{0 \leq k \leq 2} \max_{|\alpha|=k} ||\partial^{\alpha}  u_{0i} ||_{0,\gamma,\bar \Omega_s(0)}
+ \delta^*\leq 3 \delta^*. \label{P7.1-21}
\end{equation}
(iv) It follows from the Burgers' equation (\ref{PE}) that
\begin{eqnarray}
 \max_{i=1,2} |||\partial_t \hat u_i|||_{0,\gamma, \bar \Lambda_s(T;\bv)}  &\leq& 2\Big(\max_{i=1,2}|||\hat u_{i}|||_{0,\gamma, \bar \Lambda_s(T;\bv)} \Big)
  \Big( \max_{i=1,2}  \max_{|\alpha|=1} |||\partial^{\alpha} \hat u_i|||_{0,\gamma,\bar \Lambda_s(T;\bv)} \Big) \nonumber \\
  &+& |||\nabla \phi|||_{0,\gamma, \bar \Lambda_s(T;\bv)} \nonumber \\
  &\leq& 18(\delta^*)^2 + R_3. \nonumber \\
   \label{P7.1-22}
\end{eqnarray}
Finally we combine all estimates (\ref{P7.1-21}) and
(\ref{P7.1-22}) to get
\[
\max_{i=1,2} \Big( \max_{|\alpha| \leq 2} |||\partial^{\alpha}
\hat u_i|||_{0,\gamma, \bar \Lambda(T)} \Big) \leq 3 \delta^*
\quad \mbox{ and } \quad \max_{i=1, 2} |||\partial_t \hat
u_i|||_{0, \gamma, \bar \Lambda(T)} \leq \Big(18(\delta^*)^2 + R_3
\Big).
\]
By the construction of extension of $\hat \bu$, we find
\begin{eqnarray}
 && \mbox{(a)}~ \max_{i=1,2} \Big(\max_{|\alpha| \leq 2} |||\partial^{\alpha} u_i|||_{0,\gamma, \bar \Lambda(T)} \Big) \leq 3 K_0\delta^*, \label{F1} \\
 && \mbox{(b)}~ \max_{i=1, 2} |||\partial_t u_i|||_{0, \gamma, \bar \Lambda(T)} \leq K_0\Big (18(\delta^*)^2 + R_3 \Big). \label{F2}
\end{eqnarray}
Here we notice that the norm $||\cdot||_{0,\gamma,\bar
\Omega_s(t;\bv)}$ in Appendix C can be generalized to the
space-time norm $|||\cdot|||_{0,\gamma, \bar \Lambda(T)}$.
\newline

Finally the estimates (\ref{F1}) and (\ref{F2}) show that $\bu \in
{\mathcal B}(T)$.
\end{proof}
We set
\[ \Lambda_s(T; r): \mbox{ the sheath region determined by the interface $r$ },\]
and recall that an interface ${\mathcal S(t)}$ is represented by
the radial function $r(\cdot,t)$.
%
%
\begin{lemma}
\emph{(\cite{C-F})} Let $\tau < \gamma$,
\begin{eqnarray*}
&& r_i \to r \quad \mbox{ in } C^{1,\tau}(\bbr \times [0,T]) \quad
\mbox{ as given by Lemma A.5 } \quad \mbox{ and } \cr && \hat
\bu_i \in C^{1,\tau}(\Lambda_s(T;r_i)): \mbox{ be associated
solutions of (\ref{PE}) for each $i$ as given by Lemma A.10.}
\end{eqnarray*}
Let ${\mathcal E}( \hat \bu_i(\cdot,t); r_i(t))$ be the extension
of $\hat \bu_i(\cdot,t)$  with
\[  {\mathcal E}(\bu_i(\cdot,t); r_i(t)) \to \bw  \quad \mbox{ in } C^{1,\tau}(\bar \Lambda(T)). \]
Then we have
\[ \bw = {\mathcal E}(\hat \bw \Big|_{\Lambda_s(T; r)}). \]
\end{lemma}
\begin{proof}
The proof follows from a straightforward modification of the proof
in \cite{C-F} as hence is omitted.
\end{proof}

{\bf Proof of Theorem 7.3} \newline Let $\{ \bv_i\}$ be a
convergent sequence in ${\mathcal B}_T$ in the topology of
${\mathcal T}$ (see (\ref{Banach})) such that
\[ \bv_i \to \bv \quad \mbox{ in } C^{1,\tau}(\bar \Lambda(T)) \quad \mbox{ and } \quad \partial^{\alpha} \bv_i \to \partial^{\alpha} \bv,  \quad \mbox{ in }
C^{0,\tau} \quad |\alpha|=2,
 \quad 0< \tau < \gamma. \]
By Proposition A.1, ${\mathcal F}(\bv_i)$ is well-defined as an
element of ${\mathcal B}(T)$ for each $i$ and the sequence $\{
{\mathcal F}(\bv_i)\}$ is uniformly bounded in ${\mathcal T}$.
Since the Arzela-Ascoli theorem implies the compact imbedding of
$C^{1,\gamma}(\bar \Lambda(T))$ into $C^{1,\tau}(\bar \Lambda(T))$
with  $0< \tau < \gamma$, we have a convergent subsequence which
we still denote by $( \bv_i, {\mathcal F}(\bv_i))$:
\[ {\mathcal F}(\bv_i) \to \bw \qquad \mbox{ in } C^{1, \tau}(\bar \Lambda(T)). \]
We claim:
\begin{equation}
{\mathcal F}(\bv) = \bw. \label{WTS}
\end{equation}
{\it Proof of the claim}: Let $(\mbox{\boldmath $\chi$}_i, n_i,
\hat \bu_i, {\mathcal S}_i, \phi_i)$ and $(\mbox{\boldmath
$\chi$}, n, \hat \bu, {\mathcal S}, \phi)$ be the quantities
corresponding to $\bv_i$ and $\bv$ respectively.  \newline

{\it Step I}.  Suppose that
\[ \bv_i  \to \bv \mbox{ in $C^{1,\tau}(\bar \Lambda(T))$ as $i \to \infty$.} \]
Then it follows from Lemma A.1 (2) that $\mbox{\boldmath
$\chi$}_i(\cdot, t,\bx) \to \mbox{\boldmath $\chi$}(\cdot,t,\bx)
\quad $  in $C^{1,\gamma}([0,T])$ as $ i \to \infty$ and hence
since $\bv \in C^{1,\tau}(\bar \Lambda(T))$, we have
\[ \nabla \cdot \bv_i(\mbox{\boldmath $\chi$}_i(\xi, t, \bx), \xi) \to
 \nabla \cdot \bv(\mbox{\boldmath $\chi$}(\xi, t, \bx), \xi) \quad \mbox{ in }~ C^{1,\tau}(\partial \Omega_0 \times [0,T]) \qquad \mbox{ as } i \to \infty. \]
We use Lemma A.2 to get
\begin{equation}
\displaystyle \int_0^t \nabla \cdot \bv_i(\mbox{\boldmath
$\chi$}_i(\xi, t, \bx), \xi) d\xi \to \int_0^t \nabla \cdot
\bv(\mbox{\boldmath $\chi$}(\xi, t, \bx), \xi) d\xi \quad \mbox{
in } C^{1,\tau}(\partial \Omega_0 \times [0,T]) \quad \mbox{ as }
i \to \infty. \label{T7.3-1}
\end{equation}
On the other hand, since $\mbox{\boldmath $\alpha$}_i  \to
\mbox{\boldmath $\alpha$}$ in $C^{1,\tau}(\partial \Omega_0 \times
[0,T])$ as $i \to \infty$, we have
\begin{equation}
n_0(\mbox{\boldmath $\alpha$}_i(\bx,t)) \to n_0(\mbox{\boldmath
$\alpha$}(\bx,t)) \qquad \mbox{ in } C^{1,\tau}(\partial \Omega_0
\times [0,T]) \quad \mbox{ as } i \to \infty. \label{T7.3-2}
\end{equation}
Here we used the fact that $n_0$ is in $C^{1,\tau}(\bbr^2)$.
Recall the formula for $n$:
\[ n(\bx,t) =
n_0(\mbox{\boldmath $\alpha$}(\bx,t)) \exp\Big(-\int_0^t (\nabla
\cdot \bv)(\mbox{\boldmath $\chi$}(\xi,0, \mbox{\boldmath
$\alpha$}(\bx,t)), \xi)d\xi \Big). \] We now combine
(\ref{T7.3-1}) and (\ref{T7.3-2}) and the above formula to see
\[ n_i(\bx,t)  \to n(\bx,t) \qquad \mbox{ in } C^{1,\tau}(\partial \Omega_0 \times [0,T]) \quad \mbox{ as } i \to \infty, \]
which in turn implies
\[ h_{0i} = \partial_t g - (n_i \bv_i) \cdot \mbox{\boldmath $\nu$} \to h_{0} = \partial_t g - (n \bv) \cdot \mbox{\boldmath $\nu$}, \quad \mbox{ in } C^{1,\tau}(\partial \Omega_0 \times [0, T]). \]
By Lemma A.5 and Lemma A.6,  we have
\begin{eqnarray}
&& \nabla \zeta_i  \to  \nabla \zeta \quad \mbox{ in } C^{1,\tau}(\bar \Omega_* \times [0, T]) \quad \mbox{ as } i \to \infty, \label{T7.3-3} \\
&& (\theta_i, r_i, n_i) \to (\theta, r, n) \quad \mbox{ in }
C^{1,\tau}(\bbr \times [0,T]). \label{T7.3-4}
\end{eqnarray}

{\it Step II}. Let $\Lambda_s(T;\bv)$ be the sheath region
determined by $\bv$. Since ${\mathcal F}(\bv)$ is uniquely
determined by $\bv$ on the sheath region, once we can show $\bw$
satisfies equations (\ref{NEW})-(\ref{NEW1}) and the interface
conditions:
\begin{equation}
\bu = -\mbox{\boldmath $\nu$} \qquad \mbox{ and } \qquad \nabla
\phi \cdot  \mbox{\boldmath $\nu$} = 0 \qquad \mbox{ on }
{\mathcal S}(t), \label{OT}
\end{equation}
for the orthogonal flow in the sheath region $\Lambda_s(T;\bv)$,
we will have
\[ \bw = {\mathcal F}(\bv) \quad \mbox{ in } \Lambda_s(T;\bv). \]
So let us proceed to show that $\bw$ satisfies the sheath system
(\ref{NEW}) and boundary conditions (\ref{OT}) in
$\Lambda_s(T;\bv)$. Let $ {\mathcal O}$ be any open set compactly
supported in $\Lambda_s(T;\bv)$. Then by (\ref{T7.3-3}) and
(\ref{T7.3-4}), since $r_i \to r$ in $C^{1,\tau}(\bbr \times
[0,T])$, we have
 \[ {\mathcal O} \subset \Lambda_s(T; \bv_i) \quad i \geq N. \]
For $i \geq N$, we know that  $(n_i, \bv_i, \phi_i, \hat \bu_i)$
satisfy
\[
\begin{cases}
\partial_t n_i + \nabla \cdot (n_i \bv_i) = 0, \quad (\bx,t) \in {\mathcal O}, \cr
\Delta \phi_i = n_i, \cr
\partial_t \hat \bu_i + (\hat \bu_i \cdot \nabla) \hat \bu_i = \nabla \phi_i,
\end{cases}
\]
and
\[ (n_i, \bv_i, \phi_i, \hat \bu_i) \to (n, \bv, \phi, \bw) \mbox{ in } C^{1,\tau}(\bar {\mathcal O}),\]
and hence we find in the limit as $i \to \infty$,
\[
\begin{cases}
\partial_t n + \nabla \cdot (n \bv) = 0, \quad (\bx,t) \in {\mathcal O}, \cr
\Delta \phi = n, \cr
\partial_t \bw + (\bw \cdot \nabla) \bw = \nabla \phi.
\end{cases}
\]
Next we check the boundary conditions on the sheath interface.
Since by (\ref{T7.3-4}) $~~\theta_i \to \theta$ in
$C^{1,\gamma}(\bbr)$ and $\mbox{\boldmath $\nu$}_i = (\cos
\theta_i, \sin \theta_i)$, we obtain
\[ \mbox{\boldmath $\nu$}_i \to \mbox{\boldmath $\nu$}, \mbox{ in } C^{1,\gamma}(\bbr \times [0,T]). \]
On the other hand, we have
\[ \nabla \phi_i \cdot \mbox{\boldmath $\nu$}_i = 0 \quad \mbox{ and } \quad \hat \bu_i = -\mbox{\boldmath $\nu$}_i \quad \mbox{ on } {\mathcal S}_i. \]
Letting $i \to \infty$, we see
\[ \nabla \phi \cdot \mbox{\boldmath $\nu$} = 0 \quad \mbox{ and } \quad \bw = -\mbox{\boldmath $\nu$} \quad \mbox{ on } {\mathcal S}. \]
Hence we have shown that $\bw$ satisfies the sheath system
(\ref{NEW}) in the sheath region and boundary conditions
(\ref{OT}). By the uniqueness of the construction, we have
\[ {\mathcal F}(\bv) = \bw  \quad \mbox{ on } \Lambda_s(T;\bv). \]

{\it Step III}. Recall by (\ref{T7.3-4}) that we have
\[ r_i \to r \quad \mbox{ in } C^{1,\tau}(\bbr \times [0,T]) \quad \mbox{ and } \quad  \mathcal F(\bv_i) \to \bw. \]
Then by Lemma A.12, we have
\[ \bw = {\mathcal E}(\bw \Big|_{\Lambda_s(T; \bv)}) = {\mathcal F}(\bv). \]
Hence we showed that ${\mathcal F}$ is continuous in the
$C^{1,\tau}$-topology. Since ${\mathcal F}$ is a continuous map on
the compact and convex set ${\mathcal B}(T)$ of $C^{1,\gamma}$
space, by the Schauder fixed point theorem, ${\mathcal F}$ has a
fixed point  ${\bf u}$ such that
\[ {\mathcal F}(\bu) = \bu. \]
This $\bu$ is a desired smooth solution of the sheath system. This
completes the proof.

\end{subsection}
\end{section}

\newpage

%
%
\begin{section}{{\bf Gronwall-Bellman type inequality}}
In this appendix, we prove the Gronwall-Bellman type inequality.
\newline

Let $f$ be a real valued positive continuous function and suppose
a nonnegative real valued function $y$ satisfies the following
integral inequality:
\[ y(t) \leq f(t) + c^2 \int_0^t \int_0^{t_1} y(\tau)d\tau dt_1. \]
Then $y$ satisfies
\begin{eqnarray*}
y(t) &\leq& f(t) + c^2 \int_0^{t} \int_0^{t_1} f(s) \exp[c(t-2t_1
+ s)]dsdt_1
 \cr
     &=& f(t) + \Big( \max_{\tau \in [0, t]} f(\tau) \Big) {\mathcal O}(t^2), \quad \mbox{ as } t \to 0.
\end{eqnarray*}
\begin{proof}
Let us set
\[ w(t) \equiv  \int_0^t \int_0^{t_1} y(\tau)d\tau dt_1. \]
Then we have
\begin{equation}
y(t) \leq f(t) + c^2 w(t) \label{B1}.
\end{equation}
It is easy to see that
\[ w^{\prime \prime}(t) = y(t), \qquad w(0) = 0, \qquad w^{\prime}(0) = 0. \]
In (\ref{B1}), we have a differential inequality for $w$:
\[
w^{\prime \prime}(t) \leq c^2 w(t) + f(t). \label{B2}
\]
Now we introduce another dependent variable $u$ defined by
\[ w(t) =  \exp(ct) u(t). \]
By direct calculation, we obtain a differential inequality for
$u$:
\[ u^{\prime \prime} + 2 c u^{\prime} \leq f(t) \exp(-ct). \]
We multiply an integrating factor $\exp(2ct)$ to get
\[ \Big( \exp(2ct) u^{\prime}\Big)^{\prime} \leq f(t) \exp(ct). \]
Next we integrate the above inequality to get
\[ u(t) \leq \int_0^t \int_0^{t_1} f(s) \exp[-c(2t_1 -s)]dsdt_1, \]
where we used
\[ u(0) = 0, \qquad u^{\prime}(0) = 0. \]
This implies
\begin{eqnarray*}
w(t) &=& \exp(ct) u(t) \cr
     &\leq& \int_0^t \int_0^{t_1} f(s)\exp[c(t - 2t_1 + s)]dsdt_1.
\end{eqnarray*}
In (B1), we have
\begin{eqnarray*}
y(t) &\leq& f(t) + c^2 \int_0^t \int_0^{t_1} f(s)\exp[c(t - 2t_1 +
s)]dsdt_1 \cr     &\leq& f(t) + c^2 \Big( \max_{\tau \in [0,
t]}f(\tau) \Big)  \int_0^t \int_0^{t_1} \exp[c(t - 2t_1 +
s)]dsdt_1 \cr
     &=& f(t) + \Big( \max_{\tau \in [0, t]}f(\tau) \Big) {\mathcal O}(t^2) \qquad \mbox{ as } t \to 0,
\end{eqnarray*}
where we used
\[ \int_0^t \int_0^{t_1} \exp[c(t - 2t_1 + s)]dsdt_1 = \frac{1}{c^2} \Big[ -1 + \frac{1}{2} (e^{-ct} + e^{ct}) \Big]
= \frac{{\mathcal O}(t^2)}{c^2} \quad \mbox{ as } t \to 0. \]
\end{proof}
\end{section}

\newpage

%
%
\begin{section}{{\bf Extension Theorem}}
In this part, we present an extension theorem for
$C^{2,\gamma}$-functions defined on the sheath region
$\Omega_s(t;\bv), t \in [0, T]$ to the bigger domain $\Omega_1 :=
B(0,3\delta^*) - \Omega_0$. \newline

We first consider an upper bounds for the length of a convex
polygon and a simple closed convex curve inside the annulus
$A(r_1, r_2)$ defined by
\[ A(r_1, r_2) := \{ \bx \in \bbr^2: r_1 < |\bx| < r_2 \}. \]
\begin{lemma}
Let ${\mathcal P}$ and ${\mathcal C}$ be a convex $n$-polygon and
a convex curve inside the annulus $A(r_1, r_2)$ respectively. Then
we have
\[ l({\mathcal P}) \leq 2\pi r_2 \quad \mbox{ and } \quad l({\mathcal C}) \leq 2\pi r_2, \]
where $l({\mathcal P})$ and $l({\mathcal C})$ denote the lengths
of the polygon ${\mathcal P}$ and the curve ${\mathcal C}$
respectively.
\end{lemma}
\begin{proof}
(i) Let ${\mathcal P}={\mathcal P}(\bx_1, \cdots, \bx_n)$ be a
convex $n$-polygon whose vertices are $\bx_1, \cdots, \bx_n$. We
choose any point ${\bf c_0}$ inside ${\mathcal P}$, and we set
\[ \by_i:~\mbox{ the intersection point with a ray $\stackrel{\longrightarrow} {\bf c_0 x_i}$ and a circle $B(0,r_2)$}. \]
Then it is easy to see that
\begin{equation}
 l({\mathcal P}(\bx_1,\cdots, \bx_n)) \leq l({\mathcal P}(\by_1, \cdots, \by_n)). \label{B1}
\end{equation}
On the other hand we know that
\begin{equation}
 l({\mathcal P}(\by_1, \cdots, \by_n)) \leq l(B(0,r_2)) = 2\pi r_2. \label{B2}
\end{equation}
We combine (\ref{B1}) and (\ref{B2}) to obtain
\[ l({\mathcal P}) \leq 2\pi r_2. \]
(ii) Let ${\mathcal C}$ be a simple closed convex curve lying
inside $A(r_1,r_2)$. Note that for any simple closed convex curve
there exists some polygon whose sides are parts of  supporting
lines of the given convex curve. Choose a sufficiently small
positive constant $r_0 >0$. Since the curve ${\mathcal C}$ is
compact, there exists a finite open cover of ${\mathcal C}$
consisting of balls with a center $\bar \bx_i$ and a radius $r_0$,
say,
\[ \displaystyle  {\mathcal C} \subset \cup_{i=1}^{M} B(\bar \bx_i, r_0), \quad \mbox{ where } \bar \bx_i \in {\mathcal C}. \]
Consider an $M$-polygon consisting of parts of supporting lines at
$\bar \bx_i, i=1, \cdots, M$ and denote it by $\bar {\mathcal P}$.
Then it follows from the result of (i) that
\[ l({\mathcal C}) \leq l(\bar {\mathcal P}) \leq 2\pi r_2. \]
\end{proof}

Next we present the existence of a continuous linear extension
operator from $C^{1,\gamma}(\Omega_s(t;\bv))$ to
$C^{1,\gamma}(\Omega_1)$. Even though the construction of this
extension operator can be found in the literature, see for example
\cite{Adams, Evans, G-T}, we slightly modify the proofs given in
books \cite{Adams, Evans, G-T} for our purpose. \newline

{\bf The proof of Lemma A.10:} We first consider the local
extension near  one generic point on the interface and then glue
these local extensions together using the standard partition of
unity to get a global extension.  Let $t$ be given. \newline

\noindent {\it Step I} (local extension): Let $\bx_0$ be any
generic point on the interface ${\mathcal S}(t)$. Then there are
two cases: either ${\mathcal S}(t)$ is flat near $\bx_0$, lying in
the plane or it is not flat near $\bx_0$. \newline

\noindent {\it Case 1}: ${\mathcal S}(t)$ is flat near $\bx_0$
lying on some line. \newline

For simplicity, we assume $\bx_0 = (a_1, a_2)$ and the plane is
$\{ x_2 = a_2 \}$. We choose an open ball $B(\bx_0, r)$ such that
\[
\begin{cases}
B^+ := B(\bx_0,r) \cap \{ x_2 \geq a_2 \} \subset B(0, 3\delta^*)
- \Omega_s(t;\bv), \cr B^- := B(\bx_0, r) \cap \{x_2 \leq a_2\}
\subset \bar \Omega_s(t;\bv).
\end{cases}
\]
Let $f$ be any $C^{2,\gamma}$-function defined on
$\Omega_s(t;\bv)$. We extend $f$ to the ball $B^+ \cup B^-$ as
follows.
\[
\bar f(x_1,x_2) :=
\begin{cases}
6f(x_1,2a_2-x_2) - 8f(x_1, 3a_2 - 2x_2) + 3f(x_1,4a_2- 3x_2),
~~\mbox{ if } (x_1, x_2) \in B^+, \cr f(x_1,x_2), ~~\mbox{ if }
(x_1,x_2) \in B^-.
\end{cases}
\]
Notice this choice of $\bar f$ is not the same as given by Evans
\cite{Evans}, since he only desired $C^1$-regularity. We have used
a special case of the result given in \cite{Adams}.
\begin{center}
We claim: $\bar f$ is $C^{2,\gamma}$ in the ball $B$.
\end{center}
We need to show all partial derivatives are continuous at $\bx_0 =
(a_1, a_2)$. Let us write $f^- := \bar f \Big|_{B^-},~~f^+ := \bar
f|_{B^+}.$ By direct calculation we obtain
\begin{eqnarray*}
&& \bullet~~ \partial_{x_1}^{k} f^-(x_1,x_2) \cr && = 6
\partial_{x_1}^{k} f(x_1,2a_2-x_2) - 8 \partial_{x_1}^{k} f(x_1,
3a_2 - 2x_2) + 3\partial_{x_1}^{k} f(x_1,4a_2- 3x_2),~k=0,1,2, \cr
&& \bullet~~ \partial_{x_2} f^-(x_1,x_2) \cr && =  -6
\partial_{x_2} f(x_1,2a_2-x_2) + 16\partial_{x_2} f(x_1, 3a_2 -
2x_2) - 9 \partial_{x_2} f(x_1,4a_2- 3x_2), \cr &&  \bullet~~
\partial^2_{x_2} f^-(x_1,x_2) \cr && = 6 \partial_{x_2}^2
f(x_1,2a_2-x_2) - 32 \partial_{x_2}^2 f(x_1, 3a_2 - 2x_2) +
27\partial_{x_2}^2 f(x_1,4a_2- 3x_2), \cr &&  \bullet~~
\partial_{x_1}\partial_{x_2} f^-(x_1,x_2) \cr && = -6
\partial_{x_1} \partial_{x_2} f(x_1,2a_2-x_2) + 16
\partial_{x_1}\partial_{x_2} f(x_1, 3a_2 - 2x_2) - 9
\partial_{x_1}\partial_{x_2} f(x_1,4a_2- 3x_2)
\end{eqnarray*}
Now evaluate the above identities on the line $\{x_2 = a_2\}$ to
see that extended function $\bar f$ is $C^2$ in the ball $B$ and
we have
\begin{eqnarray*}
&& [\partial_{x_1}^2 f^-]_{0,\gamma,\bar B^+} \leq
31[\partial_{x_1}^2 f]_{0,\gamma,\bar B^-}, \qquad
[\partial_{x_1}\partial_{x_2} f^-]_{0,\gamma,\bar B^+} \leq
119[\partial_{x_1}\partial_{x_2} f]_{0,\gamma,\bar B^-} \cr &&
\mbox{ and } \quad [\partial_{x_2}^2 f^-]_{0,\gamma,\bar B^+} \leq
151[\partial_{x_2}^2 f]_{0,\gamma,\bar B^-}.
\end{eqnarray*}
Hence we have
\[ ||\bar f||_{2,\gamma, \bar B} \leq 151||f||_{1,\gamma,\bar B^-}. \]
\newline
\noindent {\it Case 2}: ${\mathcal S}(t)$ is not flat near
$\bx_0$. \newline

Since the interface ${\mathcal S}(t)$ is $C^{2,\gamma}$-regular,
we can find a $C^{2,\gamma}$-mapping $\Phi$ with
 inverse $\Phi^{-1}$ such that $\Phi$ straightens out ${\mathcal S}(t)$ near $\bx_0$.  We write $\by = \Phi(\bx), f^{\prime}(\by) := f(\Phi^{-1}(\by))$. We choose
 a small ball $B$ as before. Then as in {\it Case 1}, we extend $f^{\prime}$ from $B^-$ to $B$ and get
 \[ ||\bar f^{\prime} ||_{2,\gamma, \bar B}  \leq 151 ||f^{\prime}||_{2,\gamma,\bar B^-}. \]
 Let $W := \Phi^{-1}(B)$ and $W^{\pm} := \Phi^{-1}(B^{\pm})$. Then we have
 \[ ||\bar f||_{2,\gamma, \bar W} \leq 151 ||f||_{2,\gamma, \bar W^-}. \]
\newline
Now we glue local extensions together using the partition of unity
to get a global extension. \newline

\noindent {\it Step II} (Global extension): We will extend $f$
defined on $\Omega_s(t;\bv)$ to the bigger domain $\Omega_1$ such
that the extended $\bar f$ has support in $\Omega_1$. Let $r_1$ be
a sufficiently small number satisfying
\[ 0< r_1 < \min \Big \{\delta^*, 0.5 \delta_{*2} - r_b \Big \}. \]
Then for such $r_1$, we choose points $\bx_i (i=1,\cdots, M(t))$
on the curve ${\mathcal S}(t)$ such that neighboring $\bx_i$'s are
located by the part of curve with length $r$ except one pair of
points, i.e.,
\begin{eqnarray*}
 && l(\mbox{part of an interface curve connecting $\bx_i$ and $\bx_{i+1}$)} = r_0, \quad i = 1, \cdots, M(t) - 1, \cr
 && l(\mbox{part of an interface curve connecting $\bx_{M(t)}$ and $\bx_1$)} \leq r.
\end{eqnarray*}
Then the number $M(t)$ of such points are bounded by
\[ M(t) \leq \Big[ \frac{l({\mathcal S}(t))}{r_1}  \Big] + 1, \]
where the bracket is the greatest integer function. Then by Lemma
C.1, we know that
\[ M(t) \leq \Big[ \frac{4\pi \delta^*}{r_1}  \Big] + 1, \quad t \in [0, T_*]. \]
As in {\it Step I}, we extend $f$ to $B(\bx_i,r)$ for each $i$,
and denote $\bar f_i$ by the extended function. Now take an open
set $W_0$ whose closure is a compact subset of $\Omega_{s}(t)$ and
$\Omega_{s}(t) \subset  W_0(t) \cup \Big(\cup _{i=0}^{M(t)} W_i(t)
\Big)$. Let $\{\kappa_i \}$ be a partition of unity corresponding
to the open covering $\{ W_i(t)\}_{i=0}^{M(t)}$  of
$\Omega_s(t;\bv)$ and define
\[ \bar f := \sum_{i=0}^{M(t)} \kappa_i \bar f_i, \quad \bar f_0 = f. \]
It follows from {\it Step I} that
\[ ||\bar f||_{2,\gamma, \bar \Omega_1} \leq  151(M(t) + 1)||f||_{2,\gamma, \bar \Omega_s(t;\bv)}. \]
We take $K_0$ to be $151 \Big(\Big[ \frac{4\pi \delta^*}{r_1}
\Big] + 2 \Big)$ to obtain the desired result.
\end{section}

\vspace{3cm}

\noindent{\bf Acknowledgment} The research of M. Feldman was
supported by the NSF grant DMS-0200644, the research of S.Y. Ha
was supported by the grant of KRF and the research of M. Slemrod
was supported in part by the NSF grant DMS-0071463. We thank Prof.
P. Rabinowitz for pointing out to us the paper of Auchmuty and
Alexander. We also thank Profs. G. Auchmuty and S. Schochet  for
their valuable remarks. 
\end{document}